\documentclass[preprint,12pt]{elsarticle}

\usepackage{amssymb}
\usepackage[mathscr]{eucal}
\usepackage{setspace}
\usepackage{float}
\usepackage{array}
\usepackage{multirow}
\usepackage{color,soul}
\usepackage{xcolor}
\usepackage{soul}

\usepackage{amsmath}
\usepackage{geometry}
\usepackage{graphics}
\graphicspath{{Figures/}}

\usepackage{hyperref}
\hypersetup{
colorlinks   = true, 
urlcolor     = blue, 
linkcolor    = blue, 
citecolor   = blue 
}

\usepackage{lineno}

\biboptions{sort&compress}
\usepackage{blindtext}

\journal{Journal of Sound and Vibrations}

\begin{document}

\begin{frontmatter}



\title{Frequency shifts during whistling occurs as a transition between two phase synchronised limit cycles via a state of intermittency or abruptly}

\author[inst1]{Ramesh S. Bhavi}
\author[inst1]{Induja Pavithran}
\author[inst1]{R. I. Sujith}
\ead{sujith@iitm.ac.in}

\affiliation[inst1]{organization={Department of Aerospace Engineering},
        addressline={IIT Madras}, 
        city={Chennai},
        postcode={600036}, 
        country={India}}


\begin{abstract}

\textbf{Abstract:} Self-sustained oscillations arising from the interactions between the hydrodynamic and the acoustic field are disastrous in engineering systems such as segmented solid rocket motors and large gas pipelines. These self-sustained oscillations (limit cycle oscillations) are also referred to as aeroacoustic instabilities, which can be heard as a whistle. Understanding the change in dynamical state by altering the control parameter in an aeroacoustic system is critical in designing control strategies for aeroacoustic instabilities. In this study, as the control parameter Reynolds number ($Re$) is varied, we hear a change in the whistling frequency. We show that this change in frequency occurs via the state of intermittency, which has bursts of periodic fluctuations amidst the regime of the aperiodic fluctuations in acoustic pressure fluctuations. At a higher Reynolds number, we observe an abrupt transition from one limit cycle oscillation (LCO) to another limit cycle oscillation during the shift in whistling frequency. Further, we use synchronisation theory to investigate the coupled behaviour of the acoustic and the hydrodynamic fields. The acoustic pressure ($p'$) and hydrodynamic ($v'$) fluctuations during LCO exhibit phase synchronisation. Thus, we conclude that the shift in whistling frequency is a transition between the two phase synchronised limit cycle oscillations that occurs either through the state of intermittency or abruptly. The periodic bursts of intermittency correspond to the phase-synchronised periodic $p'$ and $v'$, and the aperiodic epochs correspond to the desynchronised aperiodic $p'$ and $v'$.

\end{abstract}




\end{frontmatter}



\section{Introduction}

An unsteady potential flow across a sharp-edged orifice or over a cavity generates a sound referred to as a whistle or a holetone \citep{chanaud1965some, howe1998acoustics}. The unsteady component of the potential flow acts as a source of the whistle. This whistle is manifested as self-sustained oscillations as a result of the interaction between the acoustic and the hydrodynamic fields \citep{hirschberg2004introduction}. Pleasant music from wind instruments such as flue organ pipe, recorder, flute, and human whistling result from these self-sustained oscillations \citep{fabre2012aeroacoustics,howe1975contributions}.

On the contrary, in engineering systems with confined flow through cavities, the self-sustained acoustic pressure oscillations cause fatigue and damage the structural integrity. Self-sustained oscillations in an aeroacoustic system are referred to as aeroacoustic instability. Such instabilities can arise in large solid propellant rocket motors where unburnt inhibitors used to segment grains act as orifices for the flow of hot gases \citep{nomoto1982experimental,flandro1973vortex,dunlap1981exploratory}. In large gas pipeline systems, the flow past orifices causes aeroacoustic instabilities leading to the mechanical failure of the pipe system \citep{sano2008transition}.

The holetone was first reported by \citet{sondhausstone}. Since then, various studies have conjectured the mechanism behind the holetone generation due to flow past the orifices \citep{hourigan1990aerodynamic,huang1991active,billon2005two,sano2008transition,matsuura2011direct,rayleightheoryofsound,chanaud1965some}. According to the mechanism proposed by Rayleigh \cite{rayleightheoryofsound,chanaud1965some}, there are a series of subsidiary processes, such as the origin of the shear layer instabilities in the jet, transport and magnification of these instabilities leading to the emergence of the vortices, the formation of acoustic pressure waves due to the impingement of the vortices, and the upstream transmission of the acoustic pressure waves that influence the origin of instabilities through feedback. \citet{anderson1952dependence} proposed that the periodic shedding of shear layer instabilities, separated from the leading edge of the orifice plate, produce oscillations in the effective area of the orifice. He then conjectured that these effective area oscillations lead to pressure fluctuations giving rise to the tonal sound.

\citet{hourigan1990aerodynamic} experimentally and numerically investigated the generation and the response of the acoustic waves on the shedding of vortices in a flow-through two consecutive baffles in a duct. They showed that the generation of acoustic energy depends on the phase of vortices passing the baffle and the resulting acoustic cycle. They also predicted sound sources using a model based on Howe's theory of aerodynamic sound \citep{howe1975contributions,howe1980dissipation}. According to Howe's theory, the sound power generated by vortices as they pass through an acoustic field is proportional to the scalar triple product of the acoustic particle velocity, the velocity of the vortex and the vorticity. \citet{sano2008transition}, by measuring convective velocities of the vortices, showed that during self-sustained oscillations, the shedding frequency of the vortices is locked with the acoustic modes of the duct.

Recently, researchers have investigated the dynamics of the aeroacoustic system by analysing the time series of the system under the purview of dynamical systems theory \citep{nair2016precursors,boujo2020processing,pavithran2020universality,bourquard2021whistling}. The self-sustained oscillations during whistling sound correspond to limit cycle oscillations (LCO) \cite{nair2016precursors}. The stable operation of an aeroacoustic system comprises low-magnitude aperiodic acoustic pressure oscillations. In the purview of dynamical systems theory, the dynamics of the system is usually studied by varying the control parameter to characterise the dynamical states as the state of the system changes from a steady operation to the state of aeroacoustic instability. Analyzing the unsteady acoustic pressure oscillations ($p'$) during the change from an initially normal operation to the state of aeroacoustic instability, \citet{nair2016precursors} observed that the state of intermittency presages the onset of aeroacoustic instabilities. They showed that, during the state of intermittency, the system exhibits bursts of periodic oscillations amidst the epochs of aperiodicity. Quantifying the characteristics of such an intermittent state, they provided the precursors to aeroacoustic instabilities. The state of intermittency has been reported in other studies as well, such as flow through orifices \citep{pavithran2020universality} and in grazing flows \citep{bourquard2021whistling}.

When the Reynolds number($Re$) based on the average velocity of the flow is varied in a flow through orifices, the system exhibits the variation in sound pressure level (SPL) and shifts in the dominant whistling frequency \citep{rockwell1983oscillations,huang1991active}. Several studies in the past have reported that the variation in the value of the SPL increases with $Re$ to a certain maximum value, then decreases to a lower value and again rises to approach a subsequent maximum value; this variation in SPL with $Re$ is accompanied by frequency shifts \citep{sano2008transition,tonon2011aeroacoustics,karthik2008mechanism,testud2009whistling}. However, the transition between the two whistling frequencies has not been addressed in the past literature.

In this study, as the Reynolds number increases, we report that the transition between two whistling frequencies occurs either through the state of intermittency or through an abrupt transition from one LCO to another LCO of the acoustic $p'$ fluctuations. At lower Reynolds numbers, we found that the frequency shift occurs via a state of intermittency. At higher Reynolds numbers, an abrupt jump is observed between the two LCOs during the shift in whistling frequency. 

The interaction between the acoustic and hydrodynamic fluctuations drives aeroacoustic instability. The continuous response from the acoustic field influences the size, shape, and shedding frequency of the coherent vortical structures and the shear layer instabilities \citep{huang1991active,ffowcs1989active}. Therefore, it is important to investigate the synchronisation dynamics of the sound and hydrodynamic fields to explain the dynamics observed in the aeroacoustic system. Thus, we utilise the tools of synchronisation theory to investigate the emergence of self-sustained oscillations from the interactions between the acoustic and the hydrodynamic subsystems.  

The phenomenon where the rhythm of the subsystems matches upon coupling is referred to as synchronisation. In the 17th century, Christiaan Huygens discovered the universal phenomenon of synchronisation when he observed the oscillations of the pendulum clocks hung over the wall were locked \citep{huygenslettertofather,Pikoviskysync}. In the following years, this phenomenon of synchronisation is reported in various domains, such as chemistry \citep{schreiber1982strange}, biology \citep{glass2001synchronization}, ecosystems \citep{blasius1999complex}, and engineering systems \citep{heagy1994synchronous,roy1994experimental,zdravkovich1982modification,pawar2017thermoacoustic}.

The synchronisation between the systems is generally confirmed by the locking of frequency (or phase) of the systems. The synchronisation phenomenon can be observed between periodic \citep{blekhman1995synchronization} and even chaotic subsystems \citep{boccaletti2002synchronization}. During a synchronised state of the coupled system, if the phase of the subsystems remains locked and their amplitudes remain uncorrelated, then the type of synchronisation is referred to as phase synchronisation \citep{rosenblum1996phase}. The synchronised state, where both the phases and amplitudes of the subsystems show the same behaviour, is referred to as a state of complete synchronisation \citep{Pikoviskysync}. A system of desynchronised coupled subsystems can reach the state of synchronisation by means of variation in coupling strength \citep{boccaletti2002synchronization} or feedback \citep{wang2001clustering}.

In our current study, we investigate whistling as a synchronisation phenomenon between the nonidentical subsystems, namely the acoustic and the hydrodynamic fields that the aeroacoustic system comprises. We show that the state of LCO corresponds to the phase synchronised (PS) state between the acoustic and hydrodynamic fluctuations. We also see that during the states of intermittency, the epochs of periodic bursts correspond to the phase-synchronised periodic fluctuations of acoustic $p'$ and hydrodynamic $v'$, and the aperiodic epochs correspond to the desynchronised aperiodic fluctuations of $p'$ and $v'$.

The remainder of the paper is outlined as follows. In Section \ref{sec: experimental setup}, we illustrate the details of the apparatus used in experiments and data processing. Experimental results are detailed in Section \ref{Experimental results and bifurcation diagram}. We describe the state of intermittency and LCO using the recurrence theory in Section \ref{Characterizing the dynamical states using recurrence theory}. In Section \ref{Investigating the coupled behaviour}, we study the coupled dynamics of the acoustic and hydrodynamic fields. The summary of the present study is concluded in Section \ref{Conclusions}

\section{Experimental set-up} \label{sec: experimental setup}
\begin{figure}
\centerline{\includegraphics[width=0.8\linewidth]{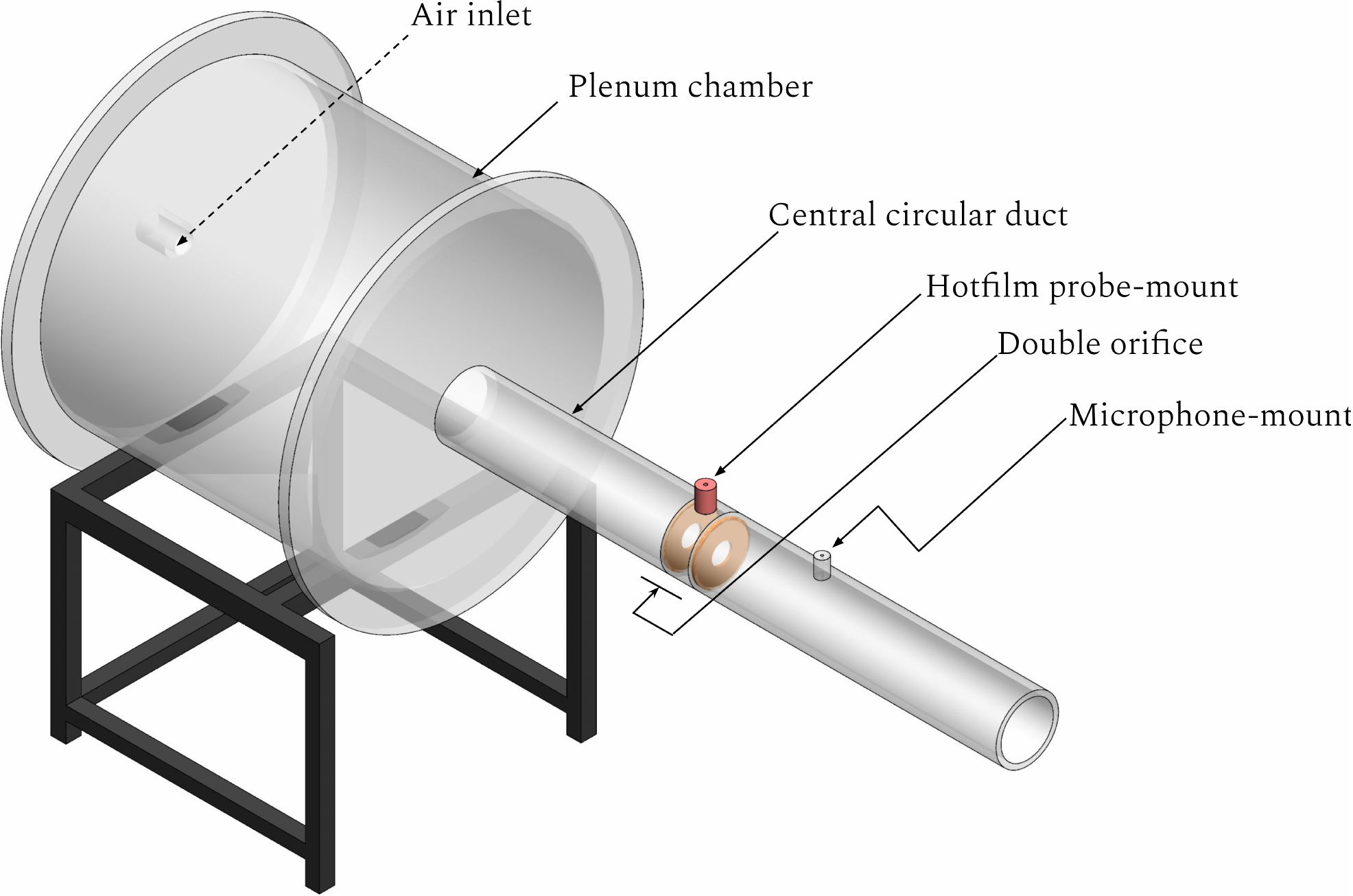}}
\caption{Schematic of the aeroacoustic system, which has a confined flow through the double orifices.} 
\label{fig: Experimental setup}
\end{figure}

The schematic of the experimental apparatus used for the study is presented in figure \ref{fig: Experimental setup}. The airflow enters the plenum chamber through an inlet port. The plenum chamber is connected to a circular central duct (610 mm length and 50 mm diameter), which has the facility to bolster two orifices (20 mm orifice diameter and 2.5 mm thickness). The distance between the orifices is 18 mm, and the first orifice, located upstream of the flow, is 220 mm away from the plenum chamber. A hot film probe, positioned between the orifices using a mount, measures hydrodynamic fluctuations. A microphone, placed on the wall of the central pipe at a distance of 305 mm from the plenum chamber, is utilized to quantify the acoustic pressure oscillations.

The control parameter of the system is the Reynolds number ($Re$), which is varied. The $Re$ is computed using the formula $Re= (\rho \Bar{v} l_c)/\mu$, where $\Bar{v}$ is the bulk flow velocity at the orifice, $\rho$ is the air density (kg/$\mathrm{m}^3$), $l_c$ is the characteristics length which is equal to the diameter of the orifice, and $\mu$ is the air- dynamic viscosity. The airflow rate is varied using an Alicat (MCR series) mass flow controller with a measurement variability of $\pm$(0.8\% of reading + 0.2\% of the complete-scale reading). The maximum error in the $Re$ is $\pm$170. The airflow rate is changed in a quasi-static way from 57 SLPM to 180 SLPM in steps of 3 SLPM. This corresponds to the variation of Reynolds number from $3956 \pm 101$ to $12632 \pm 170$. A microphone (Peizotronics PCB378C10) which has a preamplifier system and a condenser, pressure field pre-polarized, is used for measuring pressure fluctuations. The microphone has a sensitivity of 1 $\mathrm{mV}/\mathrm{Pa}$ and a 20 $\mu$Pa resolution. The data from the microphone is taken for 5 s at a sampling rate of 20 kHz. The inherent noise in the measurements from the microphone is 7 mPa.

The velocity fluctuations, $v'$, are measured using a hot film probe connected to the constant temperature anemometer (Dantec Dynamics, Multi-channel CTA-54N81). The whistling frequencies during the experiments are in the range of 400 to 550 Hz. This frequency range corresponds to the $2^{nd}$ mode of the open-open boundary condition of the duct. The microphone is mounted in the middle of the duct, near the acoustic pressure antinode. The highest acoustic velocity fluctuation at the velocity antinode is of the order of $u'_{acoustic} \thicksim \mathcal{O}(p'_{max}/\rho c)$. The RMS of the measured maximum acoustic pressure signal during whistling is 174 Pa, according to which the maximum acoustic velocity fluctuations at the velocity antinode are of the magnitude 0.35 m/s. However, the hot-film probe is placed near the node of the acoustic velocity, where the magnitude of the fluctuations will be lesser than that of the fluctuations at the antinode by order of magnitude one ($u'_{acoustic} \approx 0.035$ m/s). In this study, the value of the root mean square (RMS) velocity oscillations, measured using the hot-film probe, varies from 0.7 to 5.7 m/s. Thus, the measured fluctuations from the hot film probe represent the hydrodynamic fluctuations and are significantly higher in magnitude than the acoustic velocity fluctuations. The data from the anemometer is taken for 5 $s$ at a sampling rate of 20 kHz.

\section{Findings and discussion}\label{Sec: Results and Discussion}

\subsection{The dynamical states during the frequency shifts} \label{Experimental results and bifurcation diagram}
\begin{figure}
\includegraphics[width=\linewidth]{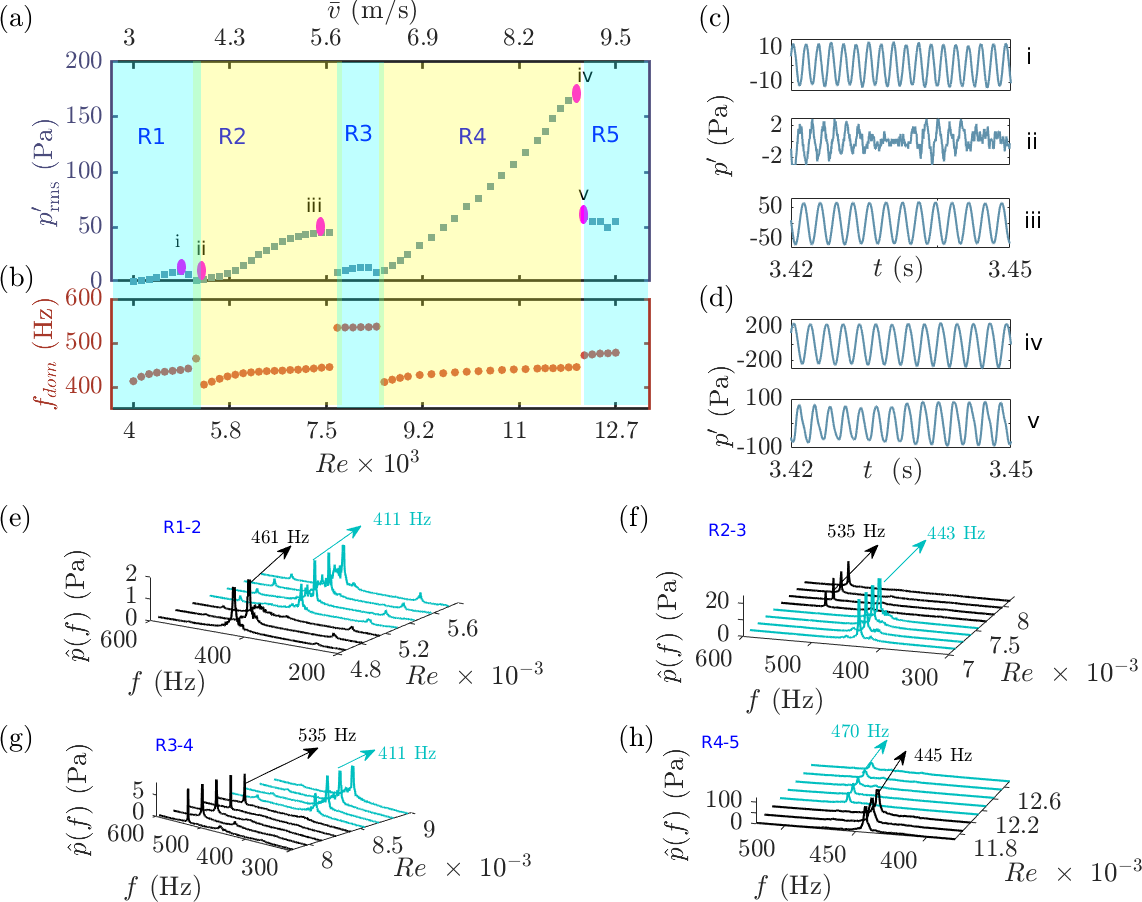}
\caption{(a) The changes in the value of the RMS of the acoustic pressure oscillations $p'_\mathrm{rms}$ (Pa) with the average flow velocity $\Bar{v}$ (m/s) of the airflow across the orifice. (b) The changes in the dominant whistling frequency of the $p'$ as the $Re$ is varied. The regions corresponding to the particular dominant frequency of $p'$ signal are categorized as R1, R2, R3, R4, and R5. The shifts in dominant frequency are observed as the $Re$  is varied. (c) The time series corresponding to the transition has the state of intermittency during the shift in whistling frequency from 461 to  411 Hz; (ci) LCO, (cii) intermittency, and (ciii) LCO. (d) Abrupt jump from one LCO (iv) to another LCO (v) during a slight shift in dominant frequency; transition from region 4-5. (e-h) Waterfall diagram of the amplitude spectrum for a particular range of Reynolds number $Re$ during the shift in dominant frequency from regions R1-2, R2-3, R3-4, and R4-5.
}
\label{Intermittency and frequency shift}
\end{figure}

To investigate the dynamical states in the aeroacoustic system having a confined flow through the double orifices, we increase the $Re$ by changing the inlet airflow rate. The $Re$ is computed based on the bulk flow velocity $\Bar{v}$ of the airflow across the orifice and the diameter $d$ of the orifice. Figure \ref{Intermittency and frequency shift}a shows the changes in the value of RMS of the acoustic pressure fluctuations $p'_\mathrm{rms}$ as $Re$ is increased. As the $Re$ increases from 4000 to 12700, we observe a rise and fall in the value of $p'_\mathrm{rms}$ reaching successive maxima followed by minima.

Figure \ref{Intermittency and frequency shift}b shows the corresponding changes in the dominant frequency (whistling frequency) of $p'$ with $Re$. We obtain the dominant frequency from the magnitude spectrum of the acoustic pressure fluctuations, computed using the fast Fourier transform (FFT). The resolution of the amplitude spectrum considered here is 0.2 Hz. We observe a switch in dominant frequencies from 461 to 411, 535 to 443, 535 to 411, and 470 to 445 Hz when the value of $Re$ crosses the values $5150$, $7500$, $8300$, and $12000$, respectively; please refer to the intersection regime of R1-2, R2-3, R3-4, and R4-5 in figure \ref{Intermittency and frequency shift}b. A detailed view of the amplitude spectrum during the frequency shifts is represented in the waterfall plots of figure \ref{Intermittency and frequency shift}e-h.

The change in the value of $p'_\mathrm{rms}$ begins from a local minimum at $Re =  4000 \pm 101$ (figure \ref{Intermittency and frequency shift}a), where the value of $p'_\mathrm{rms}$ is approximately equal to 0.7 Pa. Upon increasing the $Re$, $p'_\mathrm{rms}$ gradually increases to reach a local maximum at the value of $Re$ equal to $5150\pm 110$ (marked as i in figure \ref{Intermittency and frequency shift}a). We observe the state of LCO (figure \ref{Intermittency and frequency shift}c-i) at this local maximum ($p'_{\mathrm{rms}}=9.5$ Pa). With further increase in $Re$, the value of $p'_\mathrm{rms}$ decreases to a second local minimum ($p'_{\mathrm{rms}}=1.7 $ Pa, marked as ii in figure \ref{Intermittency and frequency shift}a). This transition from a local maximum to a local minimum is accompanied by the shift in the dominant frequency of $p'$ (figure \ref{Intermittency and frequency shift}e) from 461 to 411 Hz. We observe that this frequency switching occurs via the state of intermittency (figure \ref{Intermittency and frequency shift}c-ii), which has bursts of periodic oscillation amidst the epochs of aperiodicity. The value of $p'_\mathrm{rms}$ further increases, with increasing $Re$, to a subsequent higher local maximum ($p'_{\mathrm{rms}}=46.6$ Pa, marked as iii in figure \ref{Intermittency and frequency shift}a) when the value of $Re$ equals $7500 \pm 129$. The dynamical state corresponding to this local maximum is an LCO, as shown in figure \ref{Intermittency and frequency shift}c-iii.


Upon further increasing the Reynolds number ($Re > 7450 $), we note a sudden dip in the value of $p'_\mathrm{rms}$ from 46 to 8.5 Pa. This decrease is accompanied by a shift in dominant frequency from 443 to 535 Hz (figure \ref{Intermittency and frequency shift}f and the interface of the region R2 \& R3 in figure \ref{Intermittency and frequency shift}b). We again note that the switch in the dominant frequency occurs via the state of intermittency (refer to figure \ref{extra time series}a of \ref{Appendix section time series}). With a further increment of $Re$, we observe a similar trend of rise in $p'_{\mathrm{rms}}$, reaching a local maximum and subsequently decreasing to a minimum with a change in the dominant frequency of $p'$ fluctuations from 535 to 411 Hz. During this frequency shift from 535 to 411 Hz, we found that the transition occurs via the intermittency state (refer to figure \ref{extra time series}c of \ref{Appendix section time series}). Thus, when the value of the Reynolds number is increased from $4000 $ to $9200 $, we have shifts in whistling frequency between the regions R1-R2, R2-R3, and R3-R4 (figure \ref{Intermittency and frequency shift}b). During these frequency shifts, we observe that the dynamical state is intermittency. 

Further, an increase in the $Re$ beyond $8600$ causes $p'_{\mathrm{rms}}$ to rise again to reach a higher $p'_\mathrm{rms}$ of 170 Pa at $Re \approx 12000 \pm 165$; here we observe the dynamical state of LCO. With the continued increase in the value of $Re$, we observe an abrupt jump to another state of LCO having a lower amplitude with the value of $p'_{\mathrm{rms}}=$ 60 Pa (refer to subfigures iv \& v of figure \ref{Intermittency and frequency shift}d). This abrupt jump is accompanied by a slight frequency shift from 445 to 470 Hz. 

Thus, from figure \ref{Intermittency and frequency shift}, the aeroacoustic system under consideration in our study, which has the flow through double orifices, exhibits the state of intermittency during the frequency shifts for the control parameter $Re$ in the range of 4000 to 8600. However, for $Re$ values greater than 8600, we observe an abrupt jump from one LCO to another LCO during the frequency shift. Motivated by these findings, we utilize a visualization technique based on the theory of recurrence to characterize the dynamical states of intermittency and LCO.

\subsection{Characterizing the dynamical states using recurrence theory}\label{Characterizing the dynamical states using recurrence theory}
Recurrence is a measure attributed to the time constancy of a dynamical system \citep{Eckmannrecuurence1987}. A trajectory is said to be recurring at a location in phase space if it revisits the neighbourhood of the considered location after a specific time interval. To visualize the recurrence of the system in its phase space trajectory, a recurrence plot (RP) is built based on the recurrence matrix. In this study, we use delay embedding \citep{takens1981lecture} to compute the time-delayed vectors for the signals obtained during the experiment to build the phase space trajectory. The uniform delay embedding with time delay $\tau$ and embedding dimension $D$ is used to rebuild the trajectory of the phase space. The value of the $\tau$ is computed from the function of average mutual information (AMI) \citep{fraser1986independent}; The value of the $\tau$ at the first minimum of AMI is considered as the optimum $\tau$. The optimum dimension $D$ is computed using the false nearest neighbourhood (FNN) method \citep{kennel1992determining}. Here, the delayed vector can be written as,
\begin{equation}
\mathbf{X}_i= [ x_i,~ x_{i+\tau},~ x_{i+2\tau}...~ x_{i+(D-1)\tau}].
\end{equation}
To compute the recurrence matrix, initially, we compute the distance between the location $i$ and all the other locations of the trajectory in phase portrait. Further, we choose a distance threshold $\epsilon$ and consider only those points to compute the recurrence matrix whose Euclidean distance is lesser than the threshold $\epsilon$. The threshold $\epsilon$ can be chosen such that the number of neighbouring locations is a small part of the total span of the attractor or choose a constant number such that each point has a fixed number of neighbours \citep{marwan2011avoid}. The equation for calculating the recurrence matric $R_{ij}$ is provided as,

\begin{equation}
R_{ij} = \Theta \left ( \epsilon - \left \| \mathbf{X}_i - \mathbf{X}_j \right \| \right),
\end{equation}

Where $\Theta$ is a Heaviside step function. If the Euclidian distances $\left \| \mathbf{X}_i - \mathbf{X}_j \right \|$ is less than the threshold $\epsilon$, then the element of the matrix $R_{ij}$ is equal to one, else $R_{ij}$ is equal to zero. Entry 1 in the recurrence matrix corresponds to a recurrent state, which implies that the trajectory is revisiting its neighbourhood.

\begin{figure}
\includegraphics[width=\linewidth]{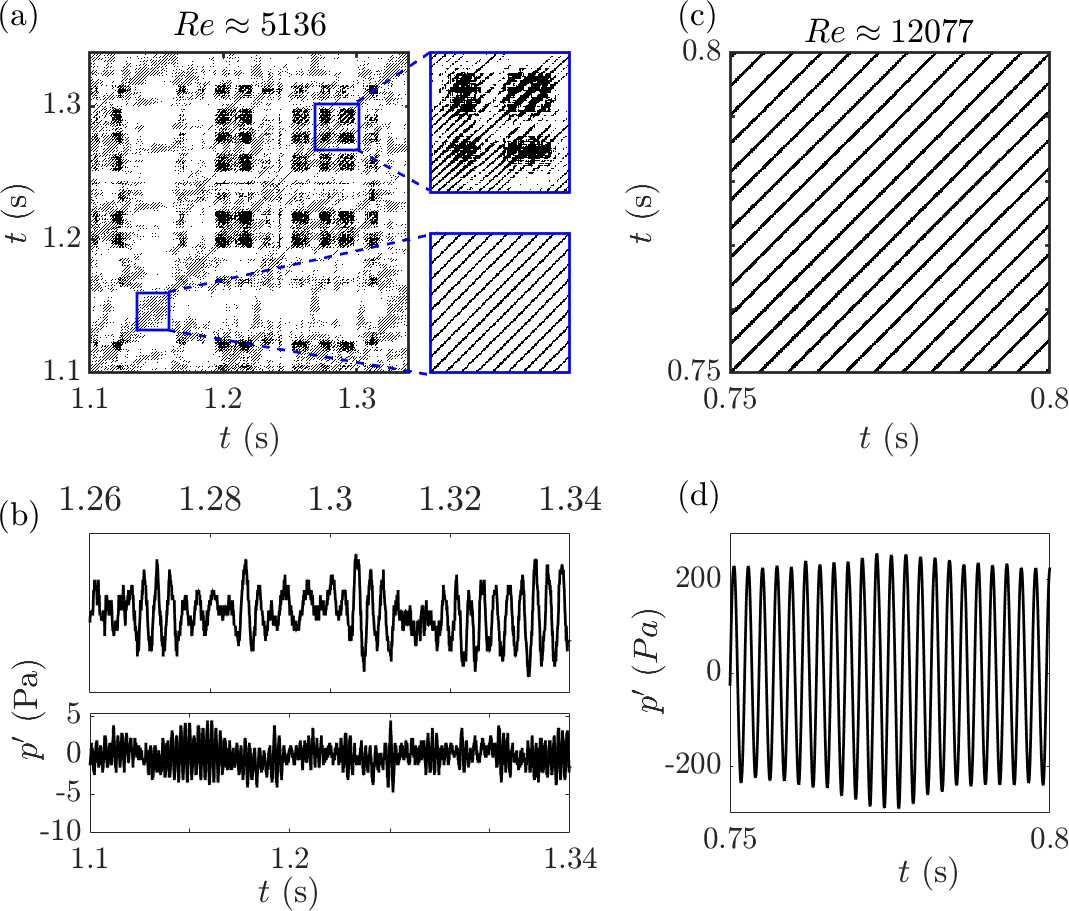}
\caption{  
    Plots for the recurrence matrix and the corresponding acoustic $p'$ fluctuations obtained during the state of (a,b) intermittency ($Re \approx 5136$) and (c,d) limit cycle oscillations ($Re \approx 12077$) of the aeroacoustic system. The recurrence plots are plotted based on choosing a fixed value for the threshold $\epsilon = \lambda/5$, where $\lambda$ is the highest distance between the pairs of points in the phase portrait. The recurrence plot for the intermittency state has black patches, which correspond to the low magnitude aperiodic oscillations relative to $\lambda$.
    }
\label{RP plots}
\end{figure}

Figure \ref{RP plots} represent the plots for the recurrence matrix obtained for the acoustic $p'$ oscillations for the state of intermittency at $Re \approx 5136$ and the state of LCO at $Re \approx 12077$. For the intermittency state, the phase space is reconstructed with an embedding dimension of $D=7$ and an optimum delay of $\tau_{opt} = 0.7$ ms; For LCO $D$ is 5 and $\tau_{opt}$ is 0.5 ms. The recurrence matrix is obtained based on choosing a fixed value for the threshold $\epsilon = \lambda/5$, where $\lambda$ is the highest span between the pairs of locations of the trajectory in phase portrait \citep{nair2014intermittency}. For the state of intermittency, the recurrence plot is seen to have perforated black patches among white regions (cf. figure \ref{RP plots}a). The occurrence of black patches in RP is due to the regime of low-magnitude aperiodic oscillations of the intermittency state. The white patches correspond to the periodic bursts of the intermittency state. Similar observations are made for the intermittency states at $Re \approx 7496$ and $Re \approx 8398$ (refer to figure \ref{Appendix figure recurrence plots} of \ref{Appendix section RP plot}). During the states of LCO, we observe equidistant diagonal lines in the RP (cf. figure \ref{RP plots}c). The time period of the LCO can be computed using the distance between the diagonal lines. The diagonal lines are also evident in the recurrence plots for the states of LCO observed at $Re \approx 4789$, 7288, 8120, and 12216 (cf. figure \ref{Appendix figure recurrence plots} of \ref{Appendix section RP plot}).

Further, the information from the topology of the recurrence plot can be quantified using the recurrence quantification measure. To obtain and compare the quantifiable measure across various values of $Re$, we fix the threshold $\epsilon$ to a specific value. 
Here we choose the threshold $\epsilon$ to be of the size of the attractor corresponding to the aperiodic state of the intermittency during frequency shift.

We make use of the measure recurrence rate ($RR$) to investigate the variation in the dynamical state of the system with Reynolds number $Re$. The $RR$ is the measure of the density of the points that recur in the RP. The equation for $RR$ is given as,
\begin{equation}\label{RR equation}
RR(\epsilon)= \frac{1}{N^2} \sum_{i,j=1}^{N}R_{i,j},(\epsilon)
\end{equation}
where $N$ is the overall number of points in the trajectory. $RR$ is the average number of black dots ($R_{ij}=1$) in the recurrence plots \citep{marwan2007recurrence}.

\begin{figure}
\includegraphics[width=\linewidth]{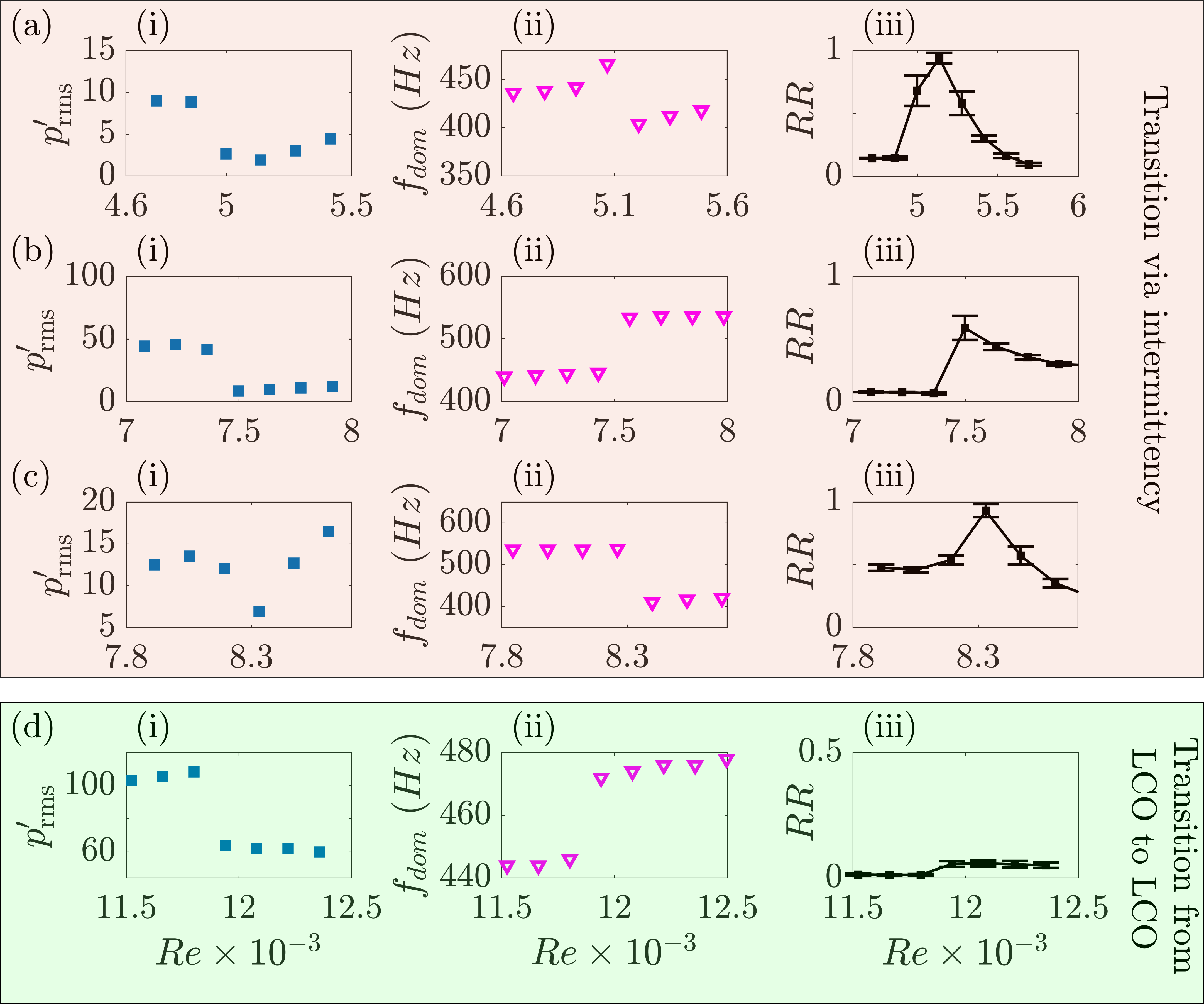}
\caption{ 
    Representation of the variation of the recurrence quantification measure during the shifts in whistling frequency via the state of intermittency (orange box) and through an abrupt jump from one LCO to another LCO (green box). (a-d)i The variation of $p'_\mathrm{rms}$ is shown in the first column for the ranges of $Re$ 4600-5500, 7000-8000, 7800-8600, and 11500-12500, respectively. (a-d)ii The corresponding variation in the dominant frequency $f$ is shown in the second column. The variation in the RQA measure (a-d)iii recurrence rate $RR$ is shown in the third column. The RQA measure is computed for an embedding dimension of 7 and the optimal $\tau$ of 0.7 ms. The signal of length 100000 points is parted into sections of 5000 points, and the mean values of RQA are plotted. The error bar represents the standard deviation. We observe a rise and dip in the value of the RQA measure, shown in the orange box, which signifies that the system has approached the state of intermittency while the whistling frequency shifts.
    }
\label{RP RQA plots}
\end{figure}

We present the variation of the recurrence rate, $RR$, during the shift in whistling frequency in figure \ref{RP RQA plots}. The first column represents the variation of $p'_\mathrm{rms}$ with $Re$ (cf. figure \ref{RP RQA plots}(a-d)-i). The second column denotes a shift in whistling frequency with $Re$ (cf. figure \ref{RP RQA plots}(a-d)-ii). The corresponding variations in $RR$ during the frequency shift are shown in the third column (cf. figure \ref{RP RQA plots}(a-d)iii). The plots corresponding to transitions via the state of intermittency are grouped in the orange background, and the abrupt transition from one LCO to another LCO is in the green background. We note that as the $Re$ varies, the curves of $RR$ increase and then decrease (cf. figure \ref{RP RQA plots}(a-c)iii). This observation can be anticipated as the number of black dots in the RP rises as the system approaches the state of intermittency as the pairwise separation length, during aperiodic epochs, now rarely cross the threshold $\epsilon$. Thus the curve of the recurrence measure $RR$ rises during the frequency shift. This variation in $RR$ confirms the presence of the intermittency state while the system transits from one whistling frequency to another. Whereas, during the abrupt transition from one LCO to another LCO in the $Re$ range of 11500 to 12500, we observe that the variation in $RR$ is negligible as both the states are of LCO and the pairwise distance regularly crosses the threshold $\epsilon$ (cf. figure \ref{RP RQA plots}ciii).

The co-existence of the acoustic and hydrodynamic subsystems in a confined flow gives rise to the synchronisation between the two. Thus, the synchronisation strength between the acoustic and hydrodynamics is an important criterion for understanding the mechanism through which intermittency (during the frequency shifts) occurs. We investigate the coupled dynamics between the acoustics and hydrodynamics of the current system using the theory of synchronisation. In this study, we cast acoustic ($p'$) and hydrodynamic ($v'$) fluctuations as two different subsystems. In the following subsection, we plot joint recurrence matrices to understand the level of synchronisation between acoustic and hydrodynamic subsystems during the LCO and the intermittency states. 

\subsection{Investigating the synchronised dynamics of the acoustic and hydrodynamic field}\label{Investigating the coupled behaviour}

The joint recurrence matrix (JRM) can be considered as an extension of the recurrence matrix to investigate the coupled dynamics of the two subsystems. A joint recurrence matrix helps visualize the recurrence of the trajectories, in the phase space, of the two subsystems at the same time \citep{goswami2013global,marwan2002nonlinear,romano2004multivariate}.

The JRM for two subsystems having the time-delayed vectors $\mathbf{X}$ and $\mathbf{Y}$ is calculated by computing the element-wise product of the individual recurrence matrices ($R^X$, $R^Y$). The equation for JRM can be written as,
\begin{equation}
JRM_{ij} = \Theta \left ( \epsilon - \left \| x_{i} - x_{j} \right \| \right)\Theta \left ( \epsilon - \left \| y_{i} - y_{j} \right \| \right).
\end{equation} 
If the trajectories $X$ and $Y$ of the two subsystems  recur simultaneously, then $JRM_{ij}=1$ else, $JRM_{ij}=0$. 


\begin{figure}
\includegraphics[width=\linewidth]{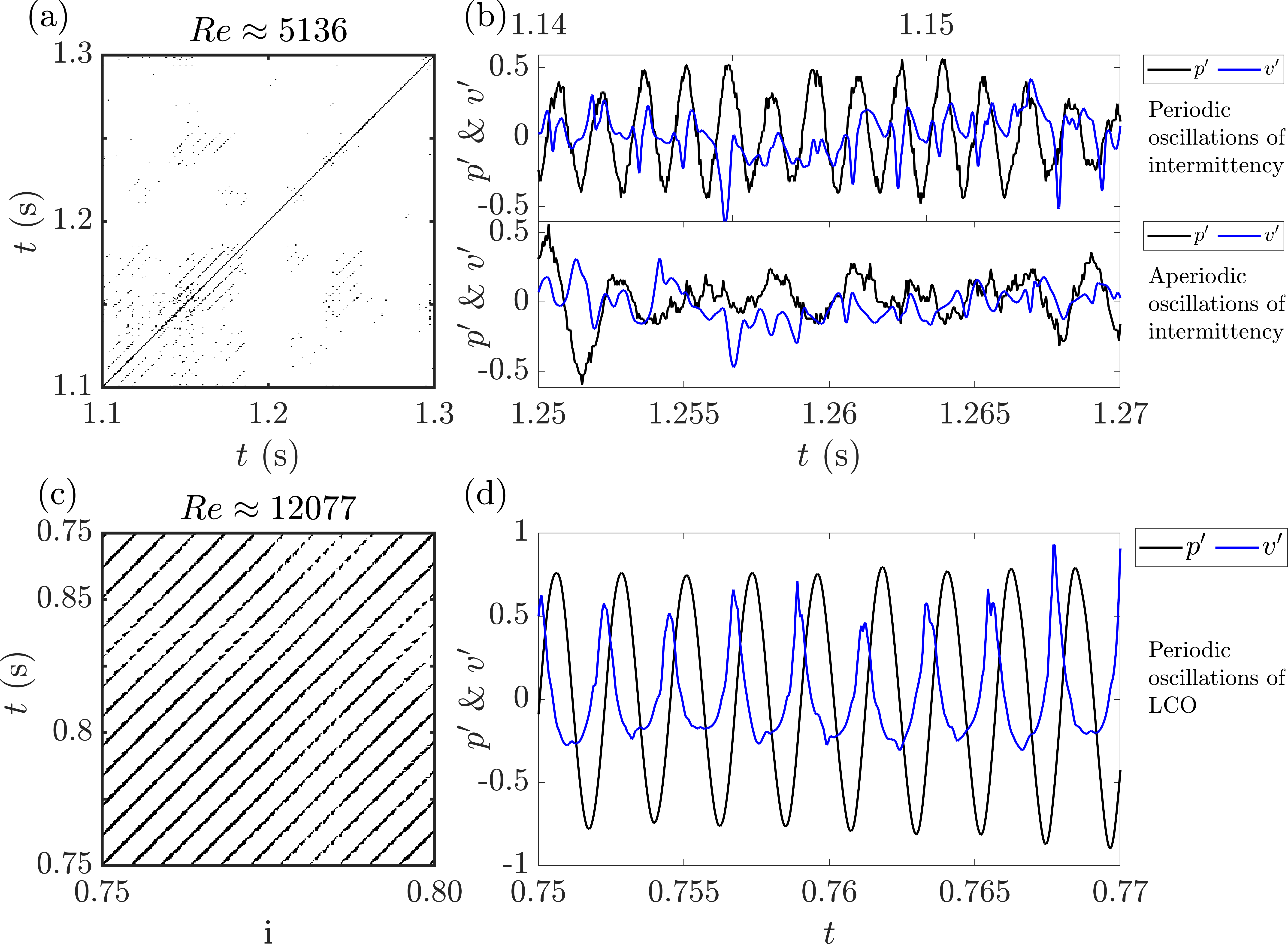}
\caption{  
The representation of the JRP for (a) the state of intermittency at $Re \approx5136$ and (c) the state of LCO at $Re \approx12077$. The corresponding normalized signal of the ($p'$) superimposed on velocity ($v'$) oscillations during the (b)intermittency state and (d) LCO are shown on the right subfigures. The presence of black dots in the joint recurrence plots represents the simultaneous recurrence of $p'$ and $v'$. Diagonal lines are observed for the simultaneous recurrence of $p'$ and $v'$; discontinuous diagonal lines are observed for the weakly coupled phase synchronised periodic limit cycle oscillations. During the periodic bursts of the intermittency state as well, we observe discontinuous diagonals in the JRP. The density of the black dots is minimum during the desynchronous state of the intermittency as a result of the very low simultaneous recurrence of $p'$ and $v'$. A fixed RR of 0.1 is chosen to compute the recurrence matrix of the individual subsystems.
}
\label{JRM plots}
\end{figure}

Figure \ref{JRM plots} represents the joint recurrence plots (JRP) of the phase trajectories of $p'$ and $v'$ corresponding to the intermittency state at $Re \approx5136$ and LCO at $Re \approx12077$ (cf. figure \ref{JRM plots}a and c). The corresponding time signal of the acoustic pressure $p'$ and velocity $v'$ fluctuations of the states of intermittency and LCO are shown in the subfigures b and d of figure \ref{JRM plots}. The $\epsilon$ is selected such that the recurrence rate (RR) for the individual recurrence matrix remains fixed, which is 0.1. A simultaneous recurrence of $p'$ and $v'$ would manifest as a black dot in the JRP. The black dots are spread in an irregular pattern during the desynchronised states of intermittency due to the aperiodic nature of the two subsystems. These black dots are sparsely distributed during the desynchronised state as a result of the fewer occurrences of the recurrences in $p'$ and $v'$, at the same time.

During the states of intermittency, we note the sparsely spaced irregular black patches due to the aperiodic epochs and the discontinuous diagonal lines due to the periodic epochs of $p'$ and $v'$ (cf. figure \ref{JRM plots}a). During the state of LCO, we observe that most of the area in JRP is filled with diagonal lines due to the periodic nature of the oscillations, as the periodic oscillations have more simultaneous recurrence than the aperiodic state. Note that the diagonal lines are more pronounced during LCO (cf. figure \ref{JRM plots}c), implying a higher correlation between $p'$ and $v'$.

We also present the JRP for the states of intermittency at $Re \approx7496$ and $Re \approx8398$ in the subfigures c and e of \ref{Appendix section JP plots}. The occurrence of diagonal lines in the JRP corresponding to the states of LCO at $Re \approx$ 4789, 7288, 8120, and 12216 also depict the higher correlation between $p'$ and $v'$. Thus, from the JRP, we observe that there are transitions from one synchronised state of LCO to the next synchronised state of LCO via a state of intermittency. We see that the strength of synchronisation is high during the states of LCO and is low during the states of intermittency.

We now quantify the topology of the joint recurrence plots using the RQA measures recurrence rate $RR_J$ and determinism $DET_J$. The $RR$ is defined in equation \ref{RR equation}. Determinism $DET$ quantifies the periodic dynamics of the system and is given as,
\begin{equation}\label{DET equation}
DET=\frac{\sum_{l=l_{min}}^{N}l F(l)}{\sum_{l=1}^{N}l F(l)},
\end{equation}
where $F(l)$ is the distribution frequency of the span of the diagonal lines in the recurrence plot. $DET$ represents the fraction of recurrence points in the RP that forms the diagonal lines. Determinism $DET_J$ in JRP measures phase synchronisation between the periodic signals.

\begin{figure}
\includegraphics[width=\linewidth]{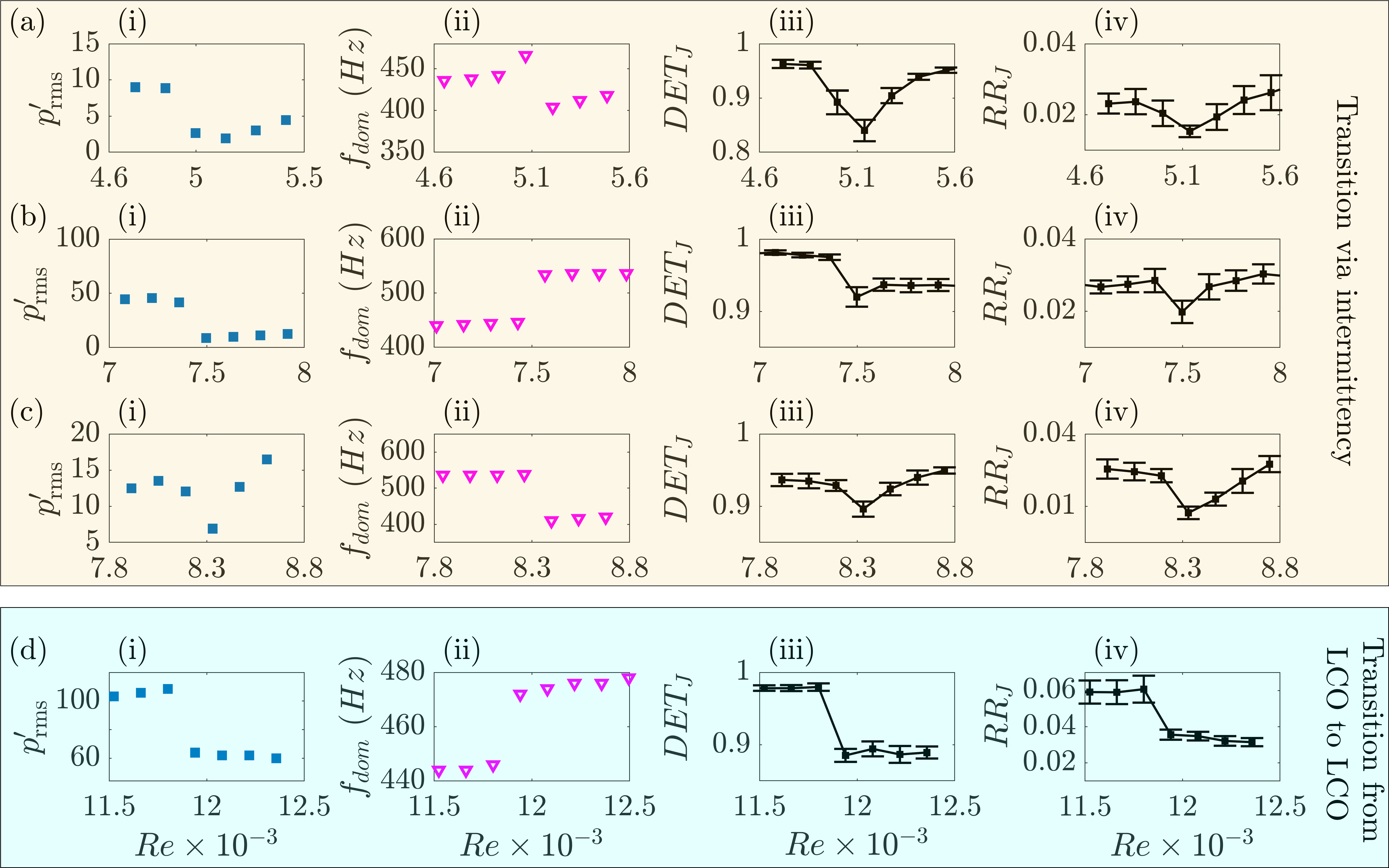}
\caption{ 
    Representation of the variation of the RQA measures for JRP during the shifts in whistling frequency via the state of intermittency (orange box) and through an abrupt jump from one LCO to another LCO (green box). (a-d)i The variation of $p'_\mathrm{rms}$ is shown in the first column for the ranges of $Re$ 4600-5500, 7000-8000, 7800-8600, and 11500-12500, respectively. (a-d)ii The corresponding variation in the dominant frequency $f_{dom}$ is shown in the second column. The variation in the RQA measures for JRP determinism (a-d)iii $DET_J$ and (a-d)iv recurrence rate $RR_J$ are shown in the third and fourth columns correspondingly. The RQA measure is calculated for an embedding dimension of 7, the optimal $\tau$ of 0.7 ms and a fixed RR of 0.1. The signal of 100000 points is parted into sections of 5000 points, and the mean values of RQA measures are plotted. The error bar represents the standard deviation. We observe a dip in the deviation of both the RQA measures, presented in the green box, during the frequency shift, which signifies a decrease in synchronisation strength during the shift in whistling frequency. 
    }
\label{JRP RQA plots}
\end{figure}

The RQA measures $RR_J$ and $DET_J$ give us information on the variation in the synchronisation strength as the Reynolds number of the system varies. Figure \ref{JRP RQA plots} represents the changes of $DET_J$ and $RR_J$ with $Re$. The first column represents the variation of $p'_\mathrm{rms}$ with $Re$ (cf. figure \ref{JRP RQA plots}(a-d)-i). The second column denotes a shift in whistling frequency with $Re$ (cf. figure \ref{JRP RQA plots}(a-d)-ii). The corresponding variations in $DET_J$ and $RR_J$ during the frequency shift are shown in the third and fourth columns, respectively (cf. figure \ref{JRP RQA plots}(a-d)iii \& iv). The plots corresponding to transitions via the state of intermittency are grouped in the orange background, and the abrupt transition from one LCO to another LCO is in the blue background. We note that as the $Re$ varies, the curves of $DET_J$ and $RR_J$ decrease and then rise, which indicates that the strength of the synchronisation reduces during the whistling frequency shift that occurs via the state of intermittency (cf. figure \ref{JRP RQA plots}(a-c)iii \& iv). There is an overall decrease in the values of $DET_J$ and $RR_J$ during the transition from high amplitude LCO to low amplitude LCO, which indicates that the synchronisation strength reduces and is manifested as the reduction in the amplitude of the LCO. Please note that the variation in these recurrence measures depends on how we define the recurrence threshold $\epsilon$ while computing the recurrence matrix. For a fixed $\epsilon$ (refer to Section \ref{Characterizing the dynamical states using recurrence theory}) we observe that the measure RR rises as the system exhibits the state of intermittency during the frequency shift. Whereas for the varying threshold, the $\epsilon$ selected such that the recurrence rate $RR$ remains comstant, we observe a dip in the value of the measure $RR_J$ as the system exhibits the state of intermittency during the frequency shift.

\begin{figure}
\includegraphics[width=\linewidth]{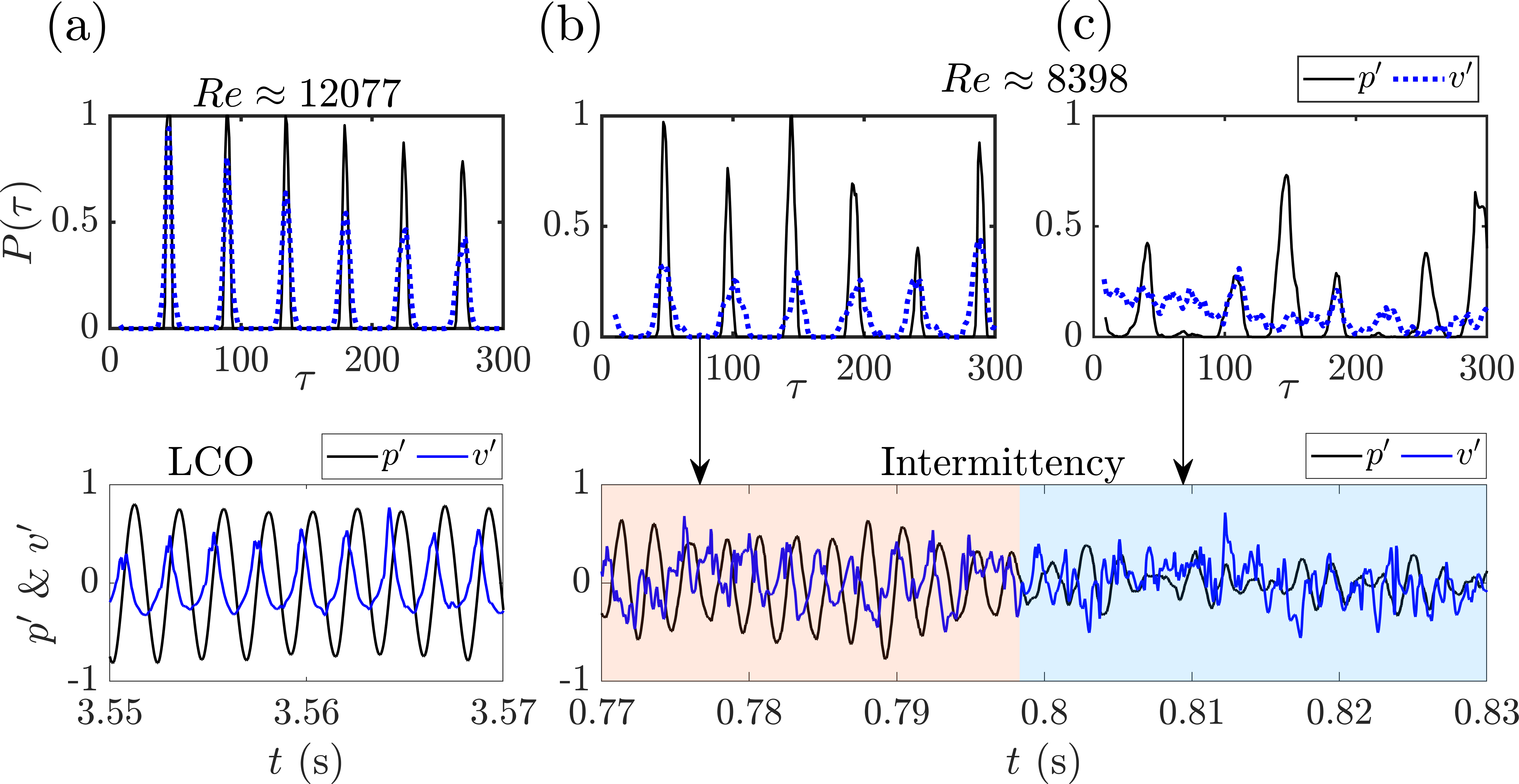}
\caption{Characterization of the type of synchronisation observed during the states of LCO and intermittency using the RPs where we observe the variation $P(\tau)$ with the time lag $\tau$. The superposition of the $P(\tau)$ curves for $p'$ and $v'$ shows (a) the state of phase synchronisation during the state of LCO, (b) the state of phase synchronisation during the periodic bursts of the intermittency, and (c) The state of desynchronisation during the aperiodic regime of the intermittency. The corresponding normalized signals of $p'$ and $v'$ are shown in the bottom row.
}
\label{Probability 
of recurrence figure}
\end{figure}

We further make use of the probability of recurrence to identify the type of synchrony that persists between $p'$ and $v'$. Probability of recurrence quantifies the probability with which a state vector of the trajectory recurs after a time lag $\tau$ \citep{romano2005detection}, and is given as,
\begin{equation}
P(\tau) = \frac{1}{N-\tau} \sum_{i=1}^{N-\tau}  \Theta \left ( \epsilon - \left \| \mathbf{X_{i}} - \mathbf{ X_{i+\tau}} \right \| \right),
\end{equation}
The recurrence property of the signal is also associated with the phase of the signal. A recurrence of the signal is equivalent to an increment in phase by $2\pi$ \citep{romano2005detection}. The synchronisation of the two coupled subsystems implies the locking of their phases and frequencies. This locking of phase leads to the simultaneous appearance of the apexes of $P(\tau)$ of two signals in the plots of probability of recurrence. If the frequencies of the two signals are locked, and their amplitude remains uncorrelated, then the state is called a phase synchronised (PS) state. The PS state manifests as the simultaneous occurrence of the peaks, but with unequal heights, in the plots of probability of recurrence with the variation in the time lag. 
On the contrary, if a functional relationship exists between the subsystems, the apexes of $P(\tau)$ for the subsystems occur simultaneously, and also, the magnitude of the peaks are matched; this state is referred to as generalized synchronisation (GS). For the generalized synchronisation state, the RP and hence the probability of recurrence plots are identical \citep{lakshmanan2011dynamics}.

In figure \ref{Probability 
of recurrence figure}, we represent the plots for variation in the probability of recurrence with a time lag $\tau$, corresponding to the state of LCO and intermittency. In order to study the coupled behaviour of the two subsystems $p'$ and $v'$, we have overlapped the $P(\tau)$ functions for $p'$ and $v'$. We observe several peaks of $P(\tau)$ at regular intervals as the time lag $\tau$ increases, denoting the existence of a very high probability of recurrence for $p'$ and $v'$ during LCO (cf. figure \ref{Probability 
of recurrence figure}a). Further, the peaks of $P(\tau)$ of the two subsystems occur simultaneously, implying that the trajectories of $p'$ and $v'$ are phase locked. Hence, we observe the state of phase synchronisation (PS) during LCO; note that the magnitude of the peaks of $P(\tau)$ for $p'$ and $v'$ are not matching, implying phase synchronisation.

During the periodic bursts of the state of intermittency, the peaks of $P(\tau)$ of the two subsystems occur simultaneously, implying that the trajectories of $p'$ and $v'$ are phase-locked. However, the amplitudes are mismatched, implying that the state is of phase synchronisation and not generalized synchronisation (cf. figure \ref{Probability 
of recurrence figure}b). During the aperiodic epochs, we have a very low $P(\tau)$ of $p'$ and $v'$, and there is no correlation among them, implying the desynchronised state  (cf. figure \ref{Probability 
of recurrence figure}c). Thus it is evident from figure \ref{Probability 
of recurrence figure}b-c that during the state of intermittency, we have the state of phase synchronisation during periodic epochs and a desynchronised state during the aperiodic regime of $p'$ and $v'$.

Thus, utilising the theory of synchronisation between acoustic and hydrodynamic fields of the aeroacoustic system, we observe a transition from one phase synchronised state (LCO) to another phase synchronised state (LCO) state as the $Re$ is increased. The shift between the phase-synchronised states happens via the intermittency state. The periodic epochs of the intermittency state have phase-synchronised periodic oscillations of pressure and hydrodynamic fluctuations. At the same time, the aperiodic epochs have desynchronised aperiodic oscillations of $p'$ and $v'$.

\section{Conclusions}\label{Conclusions}

To summarise, we presented the experimental evidence for the transitions during the shifts in whistling frequency of an aeroacoustic system having a confined flow through double orifices. The shift in whistling frequency is a transition from one limit cycle oscillation to another LCO. We showed that this transition occurs either through the state of intermittency, a state which has periodic bursts amidst the epochs of the aperiodic fluctuations, or through an abrupt transition.

We described the topology of the phase space attractor of the state of intermittency and the LCO utilising the theory of recurrence. We denote that there is a change in the dynamical state of the system during the frequency shift as the control parameter, $Re$, is varied. To show this, we quantified the graphical information of the recurrence plots using the quantification measure recurrence rate $RR$. For a fixed threshold $\epsilon$, the value of $RR$ rises during the frequency shift via the state of intermittency.

Further, we use synchronisation theory to study the coupled behaviour between the acoustic and hydrodynamic fields during the frequency shift. We found that there is a decrease in the level of synchronisation between acoustic and hydrodynamic fields during the frequency shift. We represent the level of synchronisation using the quantification measures recurrence rate $RR_J$ and determinism $DET_J$ obtained from the joint recurrence matrix. Finally, we use the probability of recurrence to investigate the type of synchronisation in the aeroacoustic system. We confirm the existence of the phase synchronisation state in our aeroacoustic system. We note that the periodic oscillations of LCO and the periodic bursts of intermittency correspond to the phase synchronised state of periodic acoustic ($p'$) and hydrodynamic ($v'$) fluctuations. The aperiodic epochs of the state of intermittency correspond to the desynchronised state of aperiodic $p'$ and $v'$.

Thus, we conclude that the shift in the whistling frequency of an aeroacoustic system is a transition between two phase synchronised limit cycle oscillations that occurs either through the state of intermittency or abruptly.

\section{Acnowledgements} \label{sec:acknowledgement}

We thank Sivakumar S. and Ankit S. for the fruitful discussions that helped us in arriving at important conclusions. We also thank Manikandan R., Anaswara B., Anand and Thilagaraj for their valuable support in experiments. We are indebted to the Department of Science and Technology, Govt. of India, for the grant under the JC Bose Fellowship (JCB/2018/000034/SSC). R.S.B. is thankful to the Ministry of Human Resource and Development, India, for granting the HTRA.


\appendix

\section{Time series of acoustic pressure during the states of LCO and intermittency}\label{Appendix section time series}
\begin{figure}
\centerline{\includegraphics[width=0.7\linewidth]{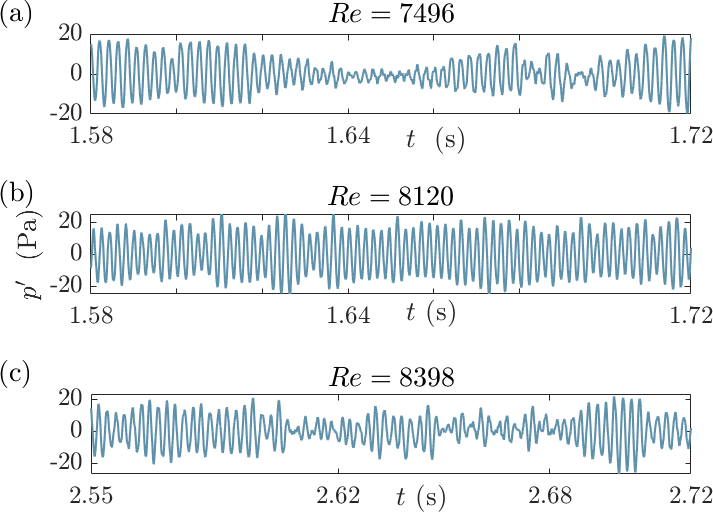}}
\caption{ Acoustic pressure fluctuations during (a) the state of intermittency at $Re \approx7496$, (b) the state of LCO at the $Re \approx8120$, and (c) the state of intermittency at $Re \approx8398$. According to the bifurcation diagram of figure \ref{Intermittency and frequency shift} in the main text, the states of intermittency are found at the minima of $p'_{rms}$ curve during the frequency shift. The maxima of $p'_{rms}$ corresponds to the state of LCO.
}
\label{extra time series}
\end{figure}

Figure \ref{extra time series} represents the time series of the acoustic pressure fluctuation signal
corresponding to the state of intermittency and LCO. The states of intermittency (\ref{extra time series}a \& c) correspond to the minima of $p'_\mathrm{rms}$ curve (during the shift whistling frequency) shown in the bifurcation diagram of figure \ref{Intermittency and frequency shift} in the main text, $Re \approx 7496~ \&~ 8398$. During this state, we observe the periodic bursts amidst the epochs of aperiodic oscillations. At $Re \approx 8120$, which corresponds to the maxima of $p'_\mathrm{rms}$ curve in figure \ref{Intermittency and frequency shift} of the main text, we observe the state of LCO.

\newpage
\section{Recurrence plots for the states of LCO and intermittency}\label{Appendix section RP plot}

\begin{figure}
\includegraphics[width=\linewidth]{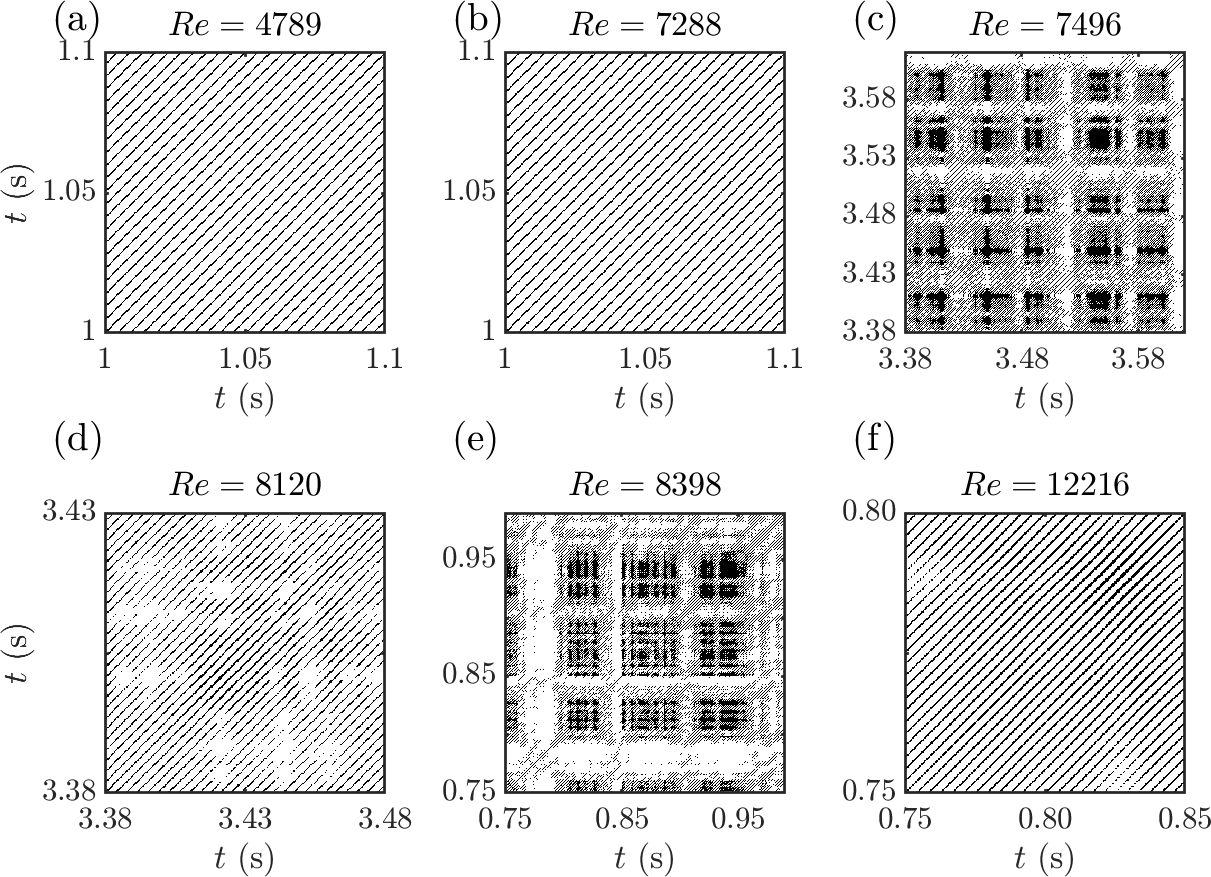}
\caption{ Recurrence plots for the dynamical states observed with variation in $Re$. (a) LCO at $Re \approx4789$. (b) LCO at $Re\approx7288$. (c) Intermittency at $Re\approx7496$. (d) LCO at $Re\approx8120$. (e) Intermittency at $Re \approx8398$. (f) LCO at $Re\approx12216$. During the states of intermittency, we observe the black patches amidst the white region of RP. Whereas, during the state of LCO, we observe equally spaced continuous diagonal lines in the RP.
}
\label{Appendix figure recurrence plots}
\end{figure}

Figure \ref{Appendix figure recurrence plots} represents the recurrence plots corresponding to the state of LCO ($Re \approx 4789, 7288, 8120 ~ \& ~ 12216$) and the state of intermittency ($Re \approx 7496 ~ \& ~ 8398$). For the state of LCO, we observe continuous diagonal lines which are equally spaced apart. The distance between the diagonal lines signifies the fundamental time period of the LCO. During the state of intermittency, we observe the black patches amidst the white region.

\newpage
\section{Joint recurrence plots for the states of LCO and intermittency}\label{Appendix section JP plots}

\begin{figure}
\includegraphics[width=\linewidth]{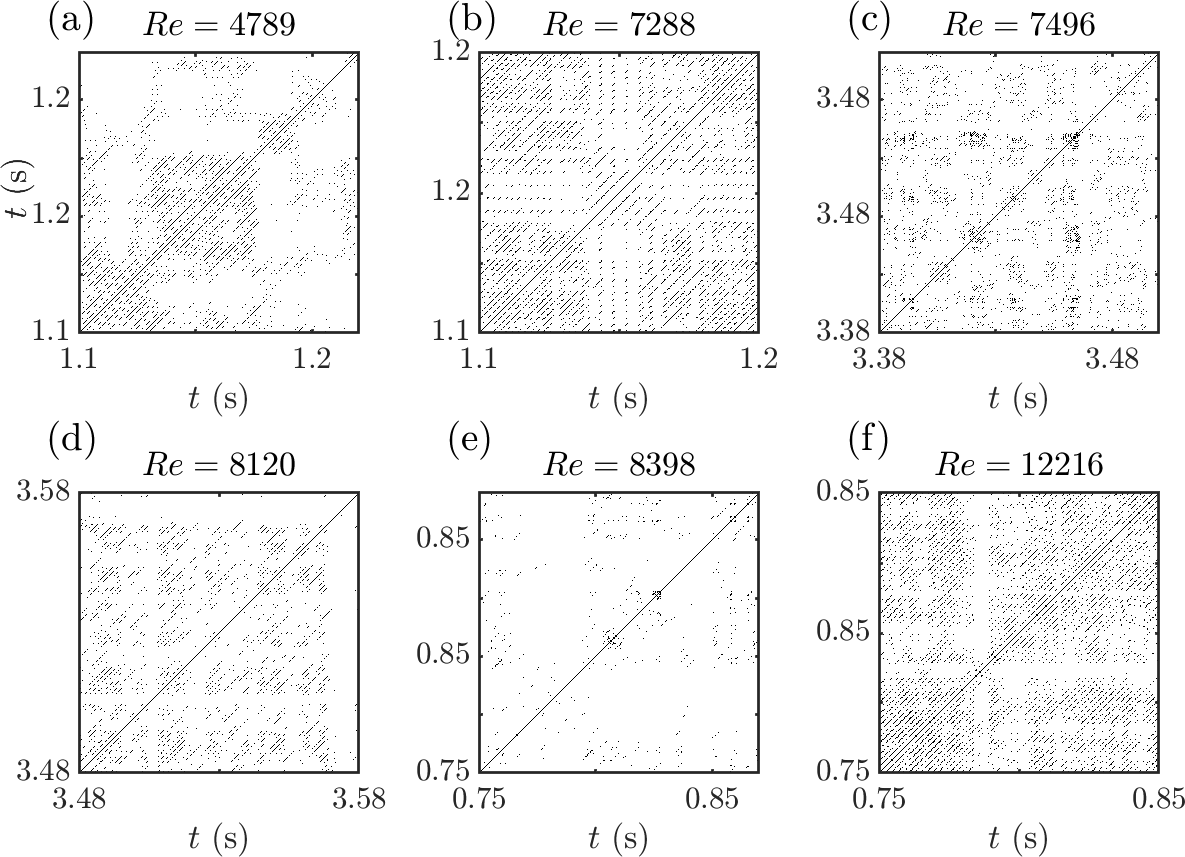}
\caption{ Joint recurrence plots of the dynamical states that are observed with variation in the $Re$. (a) LCO at $Re \approx4789$. (b) LCO at $Re \approx7288$. (c) Intermittency at $Re \approx7496$. (d) LCO at $Re \approx8120$. (e) Intermittency at $Re \approx8398$. (f) LCO at $Re \approx12216$. We observe diagonal lines in the JRP when there is higher synchronisation strength between the two periodic signals. We observe sparsely spaced black dots in JRPs when the synchronisation strength is low during the states of intermittency.
}
\label{Appendix figure joint recurrence plots}
\end{figure}

Figure \ref{Appendix figure joint recurrence plots} represents the joint recurrence plots corresponding to the state of LCO ($Re \approx 4789, 7288, 8120 ~ \& ~ 12216$) and the state of intermittency ($Re \approx 7496 ~ \& ~ 8398$). If the trajectories of the two subsystems $p'$ and $v'$ recur simultaneously, we observe a black dot in JRP. During the state of intermittency (Figure \ref{Appendix figure joint recurrence plots}c \& e), we observe that the black dots are sparsely spaced due to the absence of recurrence trajectories during the aperiodic epochs of $p'$ and $v'$. This observation depicts the weak synchronisation strength between the signals of $p'$ and $v'$. During the state of LCO, we observe an increase in the density of black dots and the appearance of short diagonal lines (Figure \ref{Appendix figure joint recurrence plots} a, b, d, and f). This observation of the diagonal lines in the JRP represents an increase in the synchronisation strength between $p'$ and $v'$.

\bibliographystyle{elsarticle-num-names} 
\bibliography{manu}

\begin{thebibliography}{51}
\expandafter\ifx\csname natexlab\endcsname\relax\def\natexlab#1{#1}\fi
\providecommand{\url}[1]{\texttt{#1}}
\providecommand{\href}[2]{#2}
\providecommand{\path}[1]{#1}
\providecommand{\DOIprefix}{doi:}
\providecommand{\ArXivprefix}{arXiv:}
\providecommand{\URLprefix}{URL: }
\providecommand{\Pubmedprefix}{pmid:}
\providecommand{\doi}[1]{\href{http://dx.doi.org/#1}{\path{#1}}}
\providecommand{\Pubmed}[1]{\href{pmid:#1}{\path{#1}}}
\providecommand{\bibinfo}[2]{#2}
\ifx\xfnm\relax \def\xfnm[#1]{\unskip,\space#1}\fi
\bibitem[{Chanaud and Powell(1965)}]{chanaud1965some}
\bibinfo{author}{R.~C. Chanaud}, \bibinfo{author}{A.~Powell},
\newblock \bibinfo{title}{Some experiments concerning the hole and ring tone},
\newblock \bibinfo{journal}{The Journal of the Acoustical Society of America}
  \bibinfo{volume}{37} (\bibinfo{year}{1965}) \bibinfo{pages}{902--911}.
\bibitem[{Howe(1998)}]{howe1998acoustics}
\bibinfo{author}{M.~S. Howe}, \bibinfo{title}{Acoustics of fluid-structure
  interactions}, \bibinfo{publisher}{Cambridge university press},
  \bibinfo{year}{1998}.
\bibitem[{Hirschberg and Rienstra(2004)}]{hirschberg2004introduction}
\bibinfo{author}{A.~Hirschberg}, \bibinfo{author}{S.~W. Rienstra},
\newblock \bibinfo{title}{An introduction to aeroacoustics},
\newblock \bibinfo{journal}{Eindhoven university of technology}
  (\bibinfo{year}{2004}).
\bibitem[{Fabre et~al.(2012)Fabre, Gilbert, Hirschberg, and
  Pelorson}]{fabre2012aeroacoustics}
\bibinfo{author}{B.~Fabre}, \bibinfo{author}{J.~Gilbert},
  \bibinfo{author}{A.~Hirschberg}, \bibinfo{author}{X.~Pelorson},
\newblock \bibinfo{title}{Aeroacoustics of musical instruments},
\newblock \bibinfo{journal}{Annual Review of Fluid Mechanics}
  \bibinfo{volume}{44} (\bibinfo{year}{2012}) \bibinfo{pages}{1--25}.
\bibitem[{Howe(1975)}]{howe1975contributions}
\bibinfo{author}{M.~S. Howe},
\newblock \bibinfo{title}{Contributions to the theory of aerodynamic sound,
  with application to excess jet noise and the theory of the flute},
\newblock \bibinfo{journal}{Journal of Fluid Mechanics} \bibinfo{volume}{71}
  (\bibinfo{year}{1975}) \bibinfo{pages}{625--673}.
\bibitem[{Nomoto and Culick(1982)}]{nomoto1982experimental}
\bibinfo{author}{H.~Nomoto}, \bibinfo{author}{F.~E.~C. Culick},
\newblock \bibinfo{title}{An experimental investigation of pure tone generation
  by vortex shedding in a duct},
\newblock \bibinfo{journal}{Journal of Sound and Vibration}
  \bibinfo{volume}{84} (\bibinfo{year}{1982}) \bibinfo{pages}{247--252}.
\bibitem[{Flandro and Jacobs(1973)}]{flandro1973vortex}
\bibinfo{author}{G.~Flandro}, \bibinfo{author}{H.~Jacobs},
\newblock \bibinfo{title}{Vortex generated sound in cavities},
\newblock in: \bibinfo{booktitle}{Aeroacoustics Conference},
  \bibinfo{year}{1973}, p. \bibinfo{pages}{1014}.
\bibitem[{Dunlap and Brown(1981)}]{dunlap1981exploratory}
\bibinfo{author}{R.~Dunlap}, \bibinfo{author}{R.~S. Brown},
\newblock \bibinfo{title}{Exploratory experiments on acoustic oscillations
  driven by periodic vortex shedding},
\newblock \bibinfo{journal}{AIAA Journal} \bibinfo{volume}{19}
  (\bibinfo{year}{1981}) \bibinfo{pages}{408--409}.
\bibitem[{Sano and Oyaizu(2008)}]{sano2008transition}
\bibinfo{author}{M.~Sano}, \bibinfo{author}{T.~Oyaizu},
\newblock \bibinfo{title}{Transition process of frequencies of pure tone caused
  by vortex shedding in a pipeline containing a double orifice},
\newblock \bibinfo{journal}{Journal of Environment and Engineering}
  \bibinfo{volume}{3} (\bibinfo{year}{2008}) \bibinfo{pages}{228--239}.
\bibitem[{Sondhauss(1854)}]{sondhausstone}
\bibinfo{author}{C.~Sondhauss},
\newblock \bibinfo{title}{Ueber die beim ausströmen der luft entstehenden
  töne},
\newblock \bibinfo{journal}{Annalen der Physik} \bibinfo{volume}{167}
  (\bibinfo{year}{1854}) \bibinfo{pages}{126--147}.
\bibitem[{Hourigan et~al.(1990)Hourigan, Welsh, Thompson, and
  Stokes}]{hourigan1990aerodynamic}
\bibinfo{author}{K.~Hourigan}, \bibinfo{author}{M.~C. Welsh},
  \bibinfo{author}{M.~C. Thompson}, \bibinfo{author}{A.~N. Stokes},
\newblock \bibinfo{title}{Aerodynamic sources of acoustic resonance in a duct
  with baffles},
\newblock \bibinfo{journal}{Journal of Fluids and Structures}
  \bibinfo{volume}{4} (\bibinfo{year}{1990}) \bibinfo{pages}{345--370}.
\bibitem[{Huang and Weaver(1991)}]{huang1991active}
\bibinfo{author}{X.~Y. Huang}, \bibinfo{author}{D.~S. Weaver},
\newblock \bibinfo{title}{On the active control of shear layer oscillations
  across a cavity in the presence of pipeline acoustic resonance},
\newblock \bibinfo{journal}{Journal of Fluids and Structures}
  \bibinfo{volume}{5} (\bibinfo{year}{1991}) \bibinfo{pages}{207--219}.
\bibitem[{Billon et~al.(2005)Billon, Valeau, and Sakout}]{billon2005two}
\bibinfo{author}{A.~Billon}, \bibinfo{author}{V.~Valeau},
  \bibinfo{author}{A.~Sakout},
\newblock \bibinfo{title}{Two feedback paths for a jet-slot oscillator},
\newblock \bibinfo{journal}{Journal of Fluids and Structures}
  \bibinfo{volume}{21} (\bibinfo{year}{2005}) \bibinfo{pages}{121--132}.
\bibitem[{Matsuura and Nakano(2011)}]{matsuura2011direct}
\bibinfo{author}{K.~Matsuura}, \bibinfo{author}{M.~Nakano},
\newblock \bibinfo{title}{Direct computation of a hole-tone feedback system at
  very low {M}ach numbers},
\newblock \bibinfo{journal}{Journal of Fluid Science and Technology}
  \bibinfo{volume}{6} (\bibinfo{year}{2011}) \bibinfo{pages}{548--561}.
\bibitem[{Rayleigh(1926)}]{rayleightheoryofsound}
\bibinfo{author}{J.~W. S.~B. Rayleigh}, \bibinfo{title}{The Theory of Sound},
  \bibinfo{publisher}{Dover Publications}, \bibinfo{year}{1926}.
\bibitem[{Anderson(1952)}]{anderson1952dependence}
\bibinfo{author}{A.~B.~C. Anderson},
\newblock \bibinfo{title}{Dependence of pfeifenton (pipe tone) frequency on
  pipe length, orifice diameter, and gas discharge pressure},
\newblock \bibinfo{journal}{The Journal of the Acoustical Society of America}
  \bibinfo{volume}{24} (\bibinfo{year}{1952}) \bibinfo{pages}{675--681}.
\bibitem[{Howe(1980)}]{howe1980dissipation}
\bibinfo{author}{M.~S. Howe},
\newblock \bibinfo{title}{The dissipation of sound at an edge},
\newblock \bibinfo{journal}{Journal of Sound and Vibration}
  \bibinfo{volume}{70} (\bibinfo{year}{1980}) \bibinfo{pages}{407--411}.
\bibitem[{Nair and Sujith(2016)}]{nair2016precursors}
\bibinfo{author}{V.~Nair}, \bibinfo{author}{R.~I. Sujith},
\newblock \bibinfo{title}{Precursors to self-sustained oscillations in
  aeroacoustic systems},
\newblock \bibinfo{journal}{International Journal of Aeroacoustics}
  \bibinfo{volume}{15} (\bibinfo{year}{2016}) \bibinfo{pages}{312--323}.
\bibitem[{Boujo et~al.(2020)Boujo, Bourquard, Xiong, and
  Noiray}]{boujo2020processing}
\bibinfo{author}{E.~Boujo}, \bibinfo{author}{C.~Bourquard},
  \bibinfo{author}{Y.~Xiong}, \bibinfo{author}{N.~Noiray},
\newblock \bibinfo{title}{Processing time-series of randomly forced
  self-oscillators: the example of beer bottle whistling},
\newblock \bibinfo{journal}{Journal of Sound and Vibration}
  \bibinfo{volume}{464} (\bibinfo{year}{2020}) \bibinfo{pages}{114--981}.
\bibitem[{Pavithran et~al.(2020)Pavithran, Unni, Varghese, Sujith, Saha,
  Marwan, and Kurths}]{pavithran2020universality}
\bibinfo{author}{I.~Pavithran}, \bibinfo{author}{V.~R. Unni},
  \bibinfo{author}{A.~J. Varghese}, \bibinfo{author}{R.~I. Sujith},
  \bibinfo{author}{A.~Saha}, \bibinfo{author}{N.~Marwan},
  \bibinfo{author}{J.~Kurths},
\newblock \bibinfo{title}{Universality in the emergence of oscillatory
  instabilities in turbulent flows},
\newblock \bibinfo{journal}{EPL (Europhysics Letters)} \bibinfo{volume}{129}
  (\bibinfo{year}{2020}) \bibinfo{pages}{24004}.
\bibitem[{Bourquard et~al.(2021)Bourquard, Faure-Beaulieu, and
  Noiray}]{bourquard2021whistling}
\bibinfo{author}{C.~Bourquard}, \bibinfo{author}{A.~Faure-Beaulieu},
  \bibinfo{author}{N.~Noiray},
\newblock \bibinfo{title}{Whistling of deep cavities subject to turbulent
  grazing flow: intermittently unstable aeroacoustic feedback},
\newblock \bibinfo{journal}{Journal of Fluid Mechanics} \bibinfo{volume}{909}
  (\bibinfo{year}{2021}) \bibinfo{pages}{A19}.
\bibitem[{Rockwell(1983)}]{rockwell1983oscillations}
\bibinfo{author}{D.~Rockwell},
\newblock \bibinfo{title}{Oscillations of impinging shear layers},
\newblock \bibinfo{journal}{AIAA journal} \bibinfo{volume}{21}
  (\bibinfo{year}{1983}) \bibinfo{pages}{645--664}.
\bibitem[{Tonon et~al.(2011)Tonon, Hirschberg, Golliard, and
  Ziada}]{tonon2011aeroacoustics}
\bibinfo{author}{D.~Tonon}, \bibinfo{author}{A.~Hirschberg},
  \bibinfo{author}{J.~Golliard}, \bibinfo{author}{S.~Ziada},
\newblock \bibinfo{title}{Aeroacoustics of pipe systems with closed branches},
\newblock \bibinfo{journal}{International Journal of Aeroacoustics}
  \bibinfo{volume}{10} (\bibinfo{year}{2011}) \bibinfo{pages}{201--275}.
\bibitem[{Karthik et~al.(2008)Karthik, Chakravarthy, and
  Sujith}]{karthik2008mechanism}
\bibinfo{author}{B.~Karthik}, \bibinfo{author}{S.~R. Chakravarthy},
  \bibinfo{author}{R.~I. Sujith},
\newblock \bibinfo{title}{Mechanism of pipe-tone excitation by flow through an
  orifice in a duct},
\newblock \bibinfo{journal}{International Journal of Aeroacoustics}
  \bibinfo{volume}{7} (\bibinfo{year}{2008}) \bibinfo{pages}{321--347}.
\bibitem[{Testud et~al.(2009)Testud, Aur{\'e}gan, Moussou, and
  Hirschberg}]{testud2009whistling}
\bibinfo{author}{P.~Testud}, \bibinfo{author}{Y.~Aur{\'e}gan},
  \bibinfo{author}{P.~Moussou}, \bibinfo{author}{A.~Hirschberg},
\newblock \bibinfo{title}{The whistling potentiality of an orifice in a
  confined flow using an energetic criterion},
\newblock \bibinfo{journal}{Journal of Sound and Vibration}
  \bibinfo{volume}{325} (\bibinfo{year}{2009}) \bibinfo{pages}{769--780}.
\bibitem[{Ffowcs and Zhao(1989)}]{ffowcs1989active}
\bibinfo{author}{J.~E. Ffowcs}, \bibinfo{author}{B.~C. Zhao},
\newblock \bibinfo{title}{The active control of vortex shedding},
\newblock \bibinfo{journal}{Journal of Fluids and Structures}
  \bibinfo{volume}{3} (\bibinfo{year}{1989}) \bibinfo{pages}{115--122}.
\bibitem[{Huygens(1665)}]{huygenslettertofather}
\bibinfo{author}{C.~Huygens},
\newblock \bibinfo{title}{Attachment to a letter to constantyn huygens, his
  father in: O.c. 5, letter no. 1335 of 26 {F}ebruary,}  (\bibinfo{year}{1665})
  \bibinfo{pages}{243--244}.
\bibitem[{Pikovsky et~al.(2003)Pikovsky, Rosenblum, and Kurths}]{Pikoviskysync}
\bibinfo{author}{A.~Pikovsky}, \bibinfo{author}{A.~Rosenblum},
  \bibinfo{author}{J.~Kurths}, \bibinfo{title}{Synchronization: A Universal
  Concept in Nonlinear Sciences}, \bibinfo{publisher}{Cambridge University
  Press}, \bibinfo{year}{2003}.
\bibitem[{Schreiber and Marek(1982)}]{schreiber1982strange}
\bibinfo{author}{I.~Schreiber}, \bibinfo{author}{M.~Marek},
\newblock \bibinfo{title}{Strange attractors in coupled reaction-diffusion
  cells},
\newblock \bibinfo{journal}{Physica D: Nonlinear Phenomena} \bibinfo{volume}{5}
  (\bibinfo{year}{1982}) \bibinfo{pages}{258--272}.
\bibitem[{Glass(2001)}]{glass2001synchronization}
\bibinfo{author}{L.~Glass},
\newblock \bibinfo{title}{Synchronization and rhythmic processes in
  physiology},
\newblock \bibinfo{journal}{Nature} \bibinfo{volume}{410}
  (\bibinfo{year}{2001}) \bibinfo{pages}{277--284}.
\bibitem[{Blasius et~al.(1999)Blasius, Huppert, and Stone}]{blasius1999complex}
\bibinfo{author}{B.~Blasius}, \bibinfo{author}{A.~Huppert},
  \bibinfo{author}{L.~Stone},
\newblock \bibinfo{title}{Complex dynamics and phase synchronization in
  spatially extended ecological systems},
\newblock \bibinfo{journal}{Nature} \bibinfo{volume}{399}
  (\bibinfo{year}{1999}) \bibinfo{pages}{354--359}.
\bibitem[{Heagy et~al.(1994)Heagy, Carroll, and Pecora}]{heagy1994synchronous}
\bibinfo{author}{J.~F. Heagy}, \bibinfo{author}{T.~L. Carroll},
  \bibinfo{author}{L.~M. Pecora},
\newblock \bibinfo{title}{Synchronous chaos in coupled oscillator systems},
\newblock \bibinfo{journal}{Physical Review E} \bibinfo{volume}{50}
  (\bibinfo{year}{1994}) \bibinfo{pages}{1874}.
\bibitem[{Roy and Thornburg~Jr(1994)}]{roy1994experimental}
\bibinfo{author}{R.~Roy}, \bibinfo{author}{K.~S. Thornburg~Jr},
\newblock \bibinfo{title}{Experimental synchronization of chaotic lasers},
\newblock \bibinfo{journal}{Physical Review Letters} \bibinfo{volume}{72}
  (\bibinfo{year}{1994}) \bibinfo{pages}{2009}.
\bibitem[{Zdravkovich(1982)}]{zdravkovich1982modification}
\bibinfo{author}{M.~M. Zdravkovich},
\newblock \bibinfo{title}{{Modification of Vortex Shedding in the
  Synchronization Range}},
\newblock \bibinfo{journal}{Journal of Fluids Engineering}
  \bibinfo{volume}{104} (\bibinfo{year}{1982}) \bibinfo{pages}{513--517}.
\bibitem[{Pawar et~al.(2017)Pawar, Seshadri, Unni, and
  Sujith}]{pawar2017thermoacoustic}
\bibinfo{author}{S.~A. Pawar}, \bibinfo{author}{A.~Seshadri},
  \bibinfo{author}{V.~R. Unni}, \bibinfo{author}{R.~I. Sujith},
\newblock \bibinfo{title}{Thermoacoustic instability as mutual synchronization
  between the acoustic field of the confinement and turbulent reactive flow},
\newblock \bibinfo{journal}{Journal of Fluid Mechanics} \bibinfo{volume}{827}
  (\bibinfo{year}{2017}) \bibinfo{pages}{664--693}.
\bibitem[{Blekhman et~al.(1995)Blekhman, Landa, and
  Rosenblum}]{blekhman1995synchronization}
\bibinfo{author}{I.~I. Blekhman}, \bibinfo{author}{P.~S. Landa},
  \bibinfo{author}{M.~G. Rosenblum},
\newblock \bibinfo{title}{Synchronization and chaotization in interacting
  dynamical systems},
\newblock \bibinfo{journal}{Applied Mechanics Reviews} \bibinfo{volume}{48}
  (\bibinfo{year}{1995}) \bibinfo{pages}{733--752}.
\bibitem[{Boccaletti et~al.(2002)Boccaletti, Kurths, Osipov, Valladares, and
  Zhou}]{boccaletti2002synchronization}
\bibinfo{author}{S.~Boccaletti}, \bibinfo{author}{J.~Kurths},
  \bibinfo{author}{G.~Osipov}, \bibinfo{author}{D.~L. Valladares},
  \bibinfo{author}{C.~S. Zhou},
\newblock \bibinfo{title}{The synchronization of chaotic systems},
\newblock \bibinfo{journal}{Physics Reports} \bibinfo{volume}{366}
  (\bibinfo{year}{2002}) \bibinfo{pages}{1--101}.
\bibitem[{Rosenblum et~al.(1996)Rosenblum, Pikovsky, and
  Kurths}]{rosenblum1996phase}
\bibinfo{author}{M.~G. Rosenblum}, \bibinfo{author}{A.~S. Pikovsky},
  \bibinfo{author}{J.~Kurths},
\newblock \bibinfo{title}{Phase synchronization of chaotic oscillators},
\newblock \bibinfo{journal}{Physical Review Letters} \bibinfo{volume}{76}
  (\bibinfo{year}{1996}) \bibinfo{pages}{1804}.
\bibitem[{Wang et~al.(2001)Wang, Kiss, and Hudson}]{wang2001clustering}
\bibinfo{author}{W.~Wang}, \bibinfo{author}{I.~Z. Kiss}, \bibinfo{author}{J.~L.
  Hudson},
\newblock \bibinfo{title}{Clustering of arrays of chaotic chemical oscillators
  by feedback and forcing},
\newblock \bibinfo{journal}{Physical Review Letters} \bibinfo{volume}{86}
  (\bibinfo{year}{2001}) \bibinfo{pages}{4954}.
\bibitem[{Eckmann et~al.(2017)Eckmann, Kamphorst, and
  Ruelle}]{Eckmannrecuurence1987}
\bibinfo{author}{J.~P. Eckmann}, \bibinfo{author}{S.~O. Kamphorst},
  \bibinfo{author}{D.~Ruelle},
\newblock \bibinfo{title}{Recurrence plots of dynamical systems},
\newblock \bibinfo{journal}{Journal of Fluid Mechanics} \bibinfo{volume}{827}
  (\bibinfo{year}{2017}) \bibinfo{pages}{664--693}.
\bibitem[{Takens(1981)}]{takens1981lecture}
\bibinfo{author}{F.~Takens},
\newblock \bibinfo{title}{Lecture notes in mathematics},
\newblock \bibinfo{journal}{by D. A. Rand and L. S. Young Springer, Berlin}
  \bibinfo{volume}{898} (\bibinfo{year}{1981}) \bibinfo{pages}{366}.
\bibitem[{Fraser and Swinney(1986)}]{fraser1986independent}
\bibinfo{author}{A.~M. Fraser}, \bibinfo{author}{H.~L. Swinney},
\newblock \bibinfo{title}{Independent coordinates for strange attractors from
  mutual information},
\newblock \bibinfo{journal}{Physical Review A} \bibinfo{volume}{33}
  (\bibinfo{year}{1986}) \bibinfo{pages}{1134}.
\bibitem[{Kennel et~al.(1992)Kennel, Brown, and
  Abarbanel}]{kennel1992determining}
\bibinfo{author}{M.~B. Kennel}, \bibinfo{author}{R.~Brown},
  \bibinfo{author}{H.~D.~I. Abarbanel},
\newblock \bibinfo{title}{Determining embedding dimension for phase-space
  reconstruction using a geometrical construction},
\newblock \bibinfo{journal}{Physical Review A} \bibinfo{volume}{45}
  (\bibinfo{year}{1992}) \bibinfo{pages}{3403}.
\bibitem[{Marwan(2011)}]{marwan2011avoid}
\bibinfo{author}{N.~Marwan},
\newblock \bibinfo{title}{How to avoid potential pitfalls in recurrence plot
  based data analysis},
\newblock \bibinfo{journal}{International Journal of Bifurcation and Chaos}
  \bibinfo{volume}{21} (\bibinfo{year}{2011}) \bibinfo{pages}{1003--1017}.
\bibitem[{Nair et~al.(2014)Nair, Thampi, and Sujith}]{nair2014intermittency}
\bibinfo{author}{V.~Nair}, \bibinfo{author}{G.~Thampi}, \bibinfo{author}{R.~I.
  Sujith},
\newblock \bibinfo{title}{Intermittency route to thermoacoustic instability in
  turbulent combustors},
\newblock \bibinfo{journal}{Journal of Fluid Mechanics} \bibinfo{volume}{756}
  (\bibinfo{year}{2014}) \bibinfo{pages}{470--487}.
\bibitem[{Marwan et~al.(2007)Marwan, R., Thiel, and
  Kurths}]{marwan2007recurrence}
\bibinfo{author}{N.~Marwan}, \bibinfo{author}{M.~C. R.},
  \bibinfo{author}{M.~Thiel}, \bibinfo{author}{J.~Kurths},
\newblock \bibinfo{title}{Recurrence plots for the analysis of complex
  systems},
\newblock \bibinfo{journal}{Physics Reports} \bibinfo{volume}{438}
  (\bibinfo{year}{2007}) \bibinfo{pages}{237--329}.
\bibitem[{Goswami et~al.(2013)Goswami, Marwan, Feulner, and
  Kurths}]{goswami2013global}
\bibinfo{author}{B.~Goswami}, \bibinfo{author}{N.~Marwan},
  \bibinfo{author}{G.~Feulner}, \bibinfo{author}{J.~Kurths},
\newblock \bibinfo{title}{How do global temperature drivers influence each
  other?},
\newblock \bibinfo{journal}{The European Physical Journal Special Topics}
  \bibinfo{volume}{222} (\bibinfo{year}{2013}) \bibinfo{pages}{861--873}.
\bibitem[{Marwan and Kurths(2002)}]{marwan2002nonlinear}
\bibinfo{author}{N.~Marwan}, \bibinfo{author}{J.~Kurths},
\newblock \bibinfo{title}{Nonlinear analysis of bivariate data with cross
  recurrence plots},
\newblock \bibinfo{journal}{Physics Letters A} \bibinfo{volume}{302}
  (\bibinfo{year}{2002}) \bibinfo{pages}{299--307}.
\bibitem[{Romano et~al.(2004)Romano, Thiel, Kurths, and von
  Bloh}]{romano2004multivariate}
\bibinfo{author}{M.~C. Romano}, \bibinfo{author}{M.~Thiel},
  \bibinfo{author}{J.~Kurths}, \bibinfo{author}{W.~von Bloh},
\newblock \bibinfo{title}{Multivariate recurrence plots},
\newblock \bibinfo{journal}{Physics Letters A} \bibinfo{volume}{330}
  (\bibinfo{year}{2004}) \bibinfo{pages}{214--223}.
\bibitem[{Romano et~al.(2005)Romano, Thiel, Kurths, Kiss, and
  Hudson}]{romano2005detection}
\bibinfo{author}{M.~C. Romano}, \bibinfo{author}{M.~Thiel},
  \bibinfo{author}{J.~Kurths}, \bibinfo{author}{I.~Z. Kiss},
  \bibinfo{author}{J.~L. Hudson},
\newblock \bibinfo{title}{Detection of synchronization for non-phase-coherent
  and non-stationary data},
\newblock \bibinfo{journal}{EPL (Europhysics Letters)} \bibinfo{volume}{71}
  (\bibinfo{year}{2005}) \bibinfo{pages}{466}.
\bibitem[{Lakshmanan and Senthilkumar(2011)}]{lakshmanan2011dynamics}
\bibinfo{author}{M.~Lakshmanan}, \bibinfo{author}{D.~V. Senthilkumar},
  \bibinfo{title}{Dynamics of nonlinear time-delay systems},
  \bibinfo{publisher}{Springer Science \& Business Media},
  \bibinfo{year}{2011}.

\end{thebibliography}





\end{document}